\documentclass[%
 preprint, 
superscriptaddress,
 amsmath,amssymb,
 aps, physrev,
]{revtex4-2}

\usepackage{amsmath}
\usepackage{amssymb}
\usepackage{xfrac}
\usepackage{graphicx}
\usepackage{caption}
\usepackage{subcaption}
\graphicspath{ {images/} }
\usepackage[dvipsnames]{xcolor}
\usepackage[margin=1in]{geometry}
\usepackage{soul}
\usepackage{comment}
\usepackage{color}
\usepackage{xcolor}
\usepackage{etoolbox}
\usepackage{parskip}
\usepackage{microtype}
\usepackage{mathtools}
\usepackage{tikz, pgfplots}
\usepackage[scr=boondox]{mathalpha} 
\usepackage{tikzscale}

\usetikzlibrary{arrows.meta}
\usetikzlibrary{positioning}
\usetikzlibrary{calc}
\usetikzlibrary{backgrounds}

\definecolor{lightblue}{rgb}{0.63, 0.74, 0.78}
\definecolor{seagreen}{rgb}{0.18, 0.42, 0.41}
\definecolor{orange}{rgb}{0.85, 0.55, 0.13}
\definecolor{silver}{rgb}{0.69, 0.67, 0.66}
\definecolor{rust}{rgb}{0.72, 0.26, 0.06}
\definecolor{purp}{RGB}{68, 14, 156}
\definecolor{joshua}{RGB}{251,220,127}
\definecolor{sky}{RGB}{86,180,233}

\colorlet{lightsilver}{silver!30!white}
\colorlet{darkorange}{orange!75!black}
\colorlet{darksilver}{silver!65!black}
\colorlet{darklightblue}{lightblue!70!black}
\colorlet{darkrust}{rust!85!black}
\colorlet{darkseagreen}{seagreen!85!black}
\definecolor{darksky}{HTML}{154c79}

\newcommand{\ens}[1]{\langle #1 \rangle}
\newcommand{\cmom}{\mathscr{c}}

\usepackage[colorlinks=true,linkcolor=darkrust,citecolor=darklightblue,urlcolor=darksilver]{hyperref}
\usepackage{cleveref}

\expandafter\def\csname ver@etex.sty\endcsname{}

\usepackage[font=small,format=plain,labelfont=bf]{caption}

\newcommand{\edit}[1]{\textcolor{black}{#1}}
\newcommand{\editt}[1]{\textcolor{black}{#1}}

\crefname{appendix}{}{}

\usepackage{fancyhdr}

\begin{document}

\preprint{APS/123-QED}

\title{\textbf{Techniques for improved statistical convergence in quantification of eddy diffusivity moments} 
}%

\author{Dana L. O.-L. Lavacot}
 \thanks{These authors contributed equally to this work (alphabetical by last name).}
\affiliation{%
  Stanford University
}%

\author{Jessie Liu}%
 \thanks{These authors contributed equally to this work (alphabetical by last name).}
\affiliation{%
  Stanford University
}%

\author{Brandon E. Morgan}
\affiliation{
 Lawrence Livermore National Laboratory
}%

\author{Ali Mani}
\affiliation{%
 Stanford University
}%

\date{\today}

\begin{abstract}
While recent approaches, such as the macroscopic forcing method (MFM) or Green's function-based approaches, can be used to compute Reynolds-averaged Navier--Stokes closure operators using forced direct numerical simulations, MFM can also be used to directly compute moments of the effective nonlocal and anisotropic eddy diffusivities. The low-order spatial and temporal moments contain limited information about the eddy diffusivity but are often sufficient for quantification and modeling of nonlocal and anisotropic effects. However, when using MFM to compute eddy diffusivity moments, the statistical convergence can be slow for higher-order moments. 
In this work, we demonstrate that using the same direct numerical simulation (DNS) for all forced MFM simulations improves statistical convergence of the eddy diffusivity moments.
We present its implementation in conjunction with a decomposition method that handles the MFM forcing semi-analytically and allows for consistent boundary condition treatment, which we develop for both scalar and momentum transport. 
We demonstrate that for a two-dimensional Rayleigh--Taylor instability case study, \edit{using the same DNS for all forced MFM simulations results in convergence with $\mathcal{O}(100)$ simulations rather than $\mathcal{O}(1000)$ simulations.} 
We then demonstrate the impacts of improved convergence on the quantification of the eddy diffusivity. 
\end{abstract}

\maketitle
\section{Introduction}
For many flows, direct numerical simulation (DNS) of the governing equations is computationally cost-prohibitive and only the averaged quantities, e.g., spatially- or temporally-averaged mean scalar or velocity fields, are needed. Reynolds-averaged Navier--Stokes (RANS) models are widely used in such applications. In RANS modeling, the flow variables are Reynolds decomposed into mean and fluctuating components, and closure models are needed for unclosed terms involving the product of fluctuating quantities, e.g., scalar fluxes or Reynolds stresses. 

Many closure models use the Boussinesq approximation \cite{boussinesq1877essai} which, under an assumption of scale separation and isotropy of the underlying mixing process, results in a local and isotropic eddy diffusivity for scalar transport or, analogously, eddy viscosity for momentum transport. For many complex flows, the Boussinesq approximation is invalid \cite{corrsin1975limitations}, and generally the eddy diffusivity is nonlocal and anisotropic \cite{romanof1985application, kraichnan1987eddy, hamba2004nonlocal}.

Recent approaches, such as the macroscopic forcing method (MFM) of \citet{mani2021macroscopic} or Green's function approach of \citet{hamba2004nonlocal, hamba2005nonlocal}, can be used to compute the nonlocal and anisotropic eddy diffusivity. These eddy diffusivities (or viscosities) are exact in that substitution of these operators back into the mean scalar (or mean momentum) equations results in DNS mean quantities. \citet{kraichnan1987eddy} derived an exact nonlocal and anisotropic expression for the scalar flux and Reynolds stress tensor using a Green's function. \citet{hamba2004nonlocal,hamba2005nonlocal} modified the expression to be feasible for numerical implementation for scalar and momentum transport, respectively. 
Because this approach needs the Green's function solution at each location in the averaged space, using a separate DNS for each location, computing the nonlocal and anisotropic eddy diffusivity requires as many DNSs as degrees of freedom in the averaged space.

\citet{mani2021macroscopic} developed MFM, a linear-algebra-based method for numerically obtaining closure operators. 
In MFM, one examines the closure operator by applying an appropriate forcing (not necessarily a Dirac delta function) to the governing equations and measures the averaged response. 
MFM can obtain the exact nonlocal and anisotropic eddy diffusivity similar to the approach of \citet{hamba2004nonlocal,hamba2005nonlocal}.
However, the primary use of MFM is to directly obtain spatial or temporal moments of the eddy diffusivity, using one simulation per desired moment.
\citet{liu2023systematic} showed how to use the limited information from a few low-order moments to model the eddy diffusivity. 
The modeled eddy diffusivity is nonlocal and matches the measured low-order moments, while the shape of its kernel approximately resembles the true kernel. 
\citet{lavacot2024} quantified nonlocal effects in two-dimensional Rayleigh--Taylor instability by using MFM-measured eddy diffusivity moments and showed improvement over a local model when including these low-order moments in the nonlocal model form suggested by \citet{liu2023systematic}.

In MFM, two sets of equations are solved: the donor and receiver equations. 
The donor equations provide flow fields needed for the solution of the receiver equations, which are macroscopically forced for probing the eddy diffusivity moments.
For example, in the case of determining the closure operator for mean scalar transport in natural convection, the donor equations would be the Navier--Stokes equations two-way-coupled with the scalar transport equation, and the receiver equations would be forced scalar transport equations with 
\edit{various} macroscopic forcings.
Each \edit{forcing is used} to obtain a specific spatiotemporal moment of the eddy diffusivity kernel.
For ease of implementation, it is natural to run simulations for each eddy diffusivity moment separately where each simulation solves its own donor and receiver equations.
However, due to numerical differences between processes in the simulations, the repeated numerical solutions of the donor equations may have differences.
The chaotic nature of turbulence can amplify these errors and lead to $\mathcal{O}(1)$ differences in the numerical solutions of the donor equations.
\edit{In this work, we examine how small differences in the donor solutions can affect the statistical error in the eddy diffusivity moments. 
Moreover, for chaotic problems with limited homogeneous dimensions, statistical convergence of the MFM-measured eddy diffusivity moments must be achieved through averaging many realizations of the flow.
We examine the importance of using identical sets of realizations, rather than randomly generating realizations, for computing eddy diffusivity moments.
As we show, differences in the donor (whether by design choice or through computational differences) can propagate errors when determining higher-order moments and lead to slow statistical convergence of the eddy diffusivity moments.
}

To avoid this slow convergence, an MFM simulation can instead use one donor to service multiple receiver equations.
That is, the individual MFM simulations described in the previous paragraph can be combined into one simulation using one donor but solving for multiple sets of receiver equations \edit{corresponding to multiple spatiotemporal moments of the eddy diffusivity}.
Using \edit{a single} donor then prevents statistical error from random processes and reduces the overall computational cost of MFM, since only one donor needs to be solved.

With a \edit{single donor approach}, we can \edit{then} also use the decomposition introduced by \citet{liu2023systematic} to semi-analytically treat the MFM forcing.
The decomposition method  (known as \textit{decomposition MFM} in this work) alleviates boundary condition issues when the MFM forcing is incompatible with the boundary conditions of the problem. \citet{liu2023systematic} demonstrated the decomposition method for scalar transport in a steady laminar inhomogeneous flow. 
In this work, we extend the decomposition method to obtain spatiotemporal moments of the eddy diffusivity for general unsteady and chaotic flows as well as momentum transport. 
We then apply decomposition MFM to scalar transport in Rayleigh-Taylor (RT) instability to illustrate the acceleration of statistical convergence due to use of a single donor and its compatibility with the proposed decomposition method.

In this work, we begin with scalar transport before generalizing to momentum transport. 
In Section \ref{sec:prob_form}, we define the nonlocal and anisotropic eddy diffusivity and its moments. 
In Section \ref{sec:MFM}, we introduce MFM for directly computing the eddy diffusivity moments.
In Section \ref{sec:deco_form}, we develop \edit{ decomposition MFM} for general unsteady and chaotic flows.
In Section \ref{sec:momentum}, we generalize to momentum transport and the nonlocal and anisotropic eddy viscosity. 
In Section \ref{sec:RTI}, as an illustrative example, we demonstrate the improved statistical convergence of the eddy diffusivity moments for 2D RT instability and discuss the modeling impacts. 

\section{Problem formulation for scalar transport}
\label{sec:prob_form}

The governing equation for a passive scalar is
\begin{align}
    \frac{\partial c}{\partial t} + \frac{\partial}{\partial x_i}\left(u_i c\right) = D_M\frac{\partial^2c}{\partial x_i \partial x_i} ,
    \label{eq:ste}
\end{align}
where $c(\mathbf{x},t)$ is a passive scalar field, $u_i(\mathbf{x},t)$ is an incompressible velocity field, and $D_M$ is the molecular diffusivity. The average, which we denote using $\ens{\cdot}$, may be defined as an ensemble average if the flow is ergodic, a temporal average if the flow is statistically stationary, and/or a spatial average if the flow is statistically homogeneous. Substitution of the Reynolds decomposition, $c=\ens{c}+c'$, into the scalar transport equation in \eqref{eq:ste} and averaging gives
\begin{align}
    \frac{\partial \ens{c}}{\partial t} + \frac{\partial}{\partial x_i}\left(\ens{u_i} \ens{c}\right) =D_M\frac{\partial^2\ens{c}}{\partial x_i \partial x_i} -\frac{\partial}{\partial x_i}\ens{u_i'c'},
\end{align}
where $\ens{u_i'c'}$ is the unclosed turbulent scalar flux that needs to be modeled.

If 1) the length and time scales of the underlying fluctuations are much smaller than those of the mean scalar gradient, and 2) the mixing by the underlying fluctuations is isotropic, then the Boussinesq approximation \cite{boussinesq1877essai} is valid, and the turbulent scalar flux can be modeled as
\begin{align}
    -\ens{u_i'c'} = D \frac{\partial \ens{c}}{\partial x_i},
\end{align}
where $D$ is a local and isotropic eddy diffusivity. Although widely-used, the Boussinesq approximation is often invalid for complex flows \cite{corrsin1975limitations}.

More generally, the turbulent scalar flux can be formulated exactly using the nonlocal and anisotropic eddy diffusivity \cite{romanof1985application, kraichnan1987eddy, hamba2004nonlocal}:
\begin{align}
    -\ens{u_i'c'}(\mathbf{x},t) = \int\int D_{ij}(\mathbf{x},\mathbf{x}',t,t') \left.\frac{\partial \ens{c}}{\partial x_j}\right|_{\mathbf{x}',t'} \text{d} \mathbf{x}' \text{d} t',
    \label{eq:gen_eddy_diff}
\end{align}
where $D_{ij}(\mathbf{x},\mathbf{x}',t,t')$ is the eddy diffusivity kernel. The eddy diffusivity is nonlocal in that the kernel allows the turbulent scalar flux to depend on the mean scalar gradient at all points in space and in its time history, $\mathbf{x}'$ and $t'$, respectively. The eddy diffusivity is  anisotropic in that the second-order tensor allows the turbulent scalar flux to depend on all directions of the mean scalar gradient. 


The eddy diffusivity may also be characterized by its moments, which are related to the eddy diffusivity kernel by considering the Taylor series expansion locally about $\mathbf{x}' = \mathbf{x}$ and $t' = t$ (also known as a Kramers--Moyal expansion \cite{kampen1961power}):

\begin{align} 
    -\ens{u_i'c'}(\mathbf{x},t) = \int\int D_{ij}(\mathbf{x},\mathbf{x}',t,t') 
    \left(
    \left.\frac{\partial \ens{c}}{\partial x_j}\right|_{\mathbf{x},t} + 
    \left.\left(x'_k-x_k\right)\frac{\partial^2 \ens{c}}{\partial x_k\partial x_j}\right|_{\mathbf{x},t} \right.\nonumber\\\left.
    + 
    \dots + \left.\left(t'-t\right)\frac{\partial^2 \ens{c}}{\partial t\partial x_j}\right|_{\mathbf{x},t} +
    \dots
    \right)
    \text{d}\mathbf{x}' \text{d}t'    
    \label{eq:taylor_exp}
\end{align}
Since the derivatives of $\ens{c}$ are no longer functions of $\mathbf{x}'$ and $t'$, they can be moved out of the integral, and the above equation can be rearranged as
\begin{align}
    -\ens{u_i'c'}(\textbf{x},t) = \left[D_{ij}^{00}(\mathbf{x},t) +
    D_{ijk}^{10}(\mathbf{x},t)\frac{\partial}{\partial x_k} + 
    \dots + 
    D_{ij}^{01}(\mathbf{x},t)\frac{\partial}{\partial t} +
    \dots\right] \frac{\partial \ens{c}}{\partial x_j},  
    \label{eq:kramers-moyal}
\end{align}
where the eddy diffusivity moments, $D^{mn}_{ij\dots}(\mathbf{x},t)$ are defined as
\begin{align}
    \label{eq:D00}
    D^{00}_{ij}(\mathbf{x},t) &= \int \int D_{ij}(\mathbf{x},\mathbf{x}',t,t')\text{d}\mathbf{x}' \text{d}t', 
    \\
    \label{eq:D10}
    D^{10}_{ijk}(\mathbf{x},t) &= \int \int (x_k'-x_k)D_{ij}(\mathbf{x},\mathbf{x}',t,t')\text{d}\textbf{x}' \text{d}t',
    \\
    &\vdots \nonumber
    \\
    \label{eq:D01}
    D^{01}_{ij}(\mathbf{x},t) &= \int \int (t'-t)D_{ij}(\mathbf{x},\mathbf{x}',t,t')\text{d}\mathbf{x}' \text{d}t',
    \\
    &\vdots \nonumber
\end{align}
The superscripts denote the $m$-th order spatial moment and $n$-th order temporal moment.
The leading-order term in the expansion is the zeroth-order spatiotemporal moment, $D^{00}_{ij}$, and is local and anisotropic. 
The higher-order spatiotemporal moments can be used to characterize nonlocal effects. 
For example, by using MFM \cite{mani2021macroscopic} to measure the moments, \citet{park2024direct} investigated nonlocality and anisotropy in turbulent channel flow and \citet{lavacot2024} investigated spatiotemporal nonlocality in Rayleigh--Taylor instability. \citet{liu2023systematic} showed how to use the low-order eddy diffusivity moments to model the nonlocal eddy diffusivity kernel.

\section{Eddy diffusivity moments using the macroscopic forcing method (MFM)}
\label{sec:MFM}
In this section, we briefly introduce MFM for computing closure operators, e.g., the nonlocal and anisotropic eddy diffusivity in Equation \eqref{eq:gen_eddy_diff}, before showing MFM for directly computing the eddy diffusivity moments in Equations \eqref{eq:D00}-\eqref{eq:D01}.
We discuss slow convergence in the higher-order moments due to error propagation from the lower-order moments. 
 
In MFM, forcing is added to the scalar transport equation:
\begin{align}
    \frac{\partial c}{\partial t} + \frac{\partial}{\partial x_i}\left(u_i c\right) = D_M\frac{\partial^2c}{\partial x_i \partial x_i} + s,
    \label{eq:forced_ste}
\end{align}
where $s$ is the MFM forcing with the important macroscopic property, $s = \ens{s}$. As detailed in \citet{mani2021macroscopic}, by explicitly specifying the MFM forcing, one can arrive at the closure operator by post-processing $\ens{c}$. Alternatively, by using the MFM forcing to maintain a specified mean scalar, $ \ens{c}$, one can also arrive at the closure operator using what is known as inverse MFM. For example, one can compute the nonlocal and anisotropic eddy diffusivity in \eqref{eq:gen_eddy_diff} by specifying the mean scalar such that the gradient is a Dirac delta function at each point in the averaged space and post-processing the turbulent scalar flux. Each point requires a separate forced DNS, and thus obtaining the eddy diffusivity kernel for the entire domain requires as many DNSs as degrees of freedom in the averaged space. Due to the large number of DNSs needed, this brute force approach is computationally expensive and practically infeasible for problems with many degrees of freedom in the averaged space. The brute force application of inverse MFM with Dirac delta functions results in an approach identical to the Green's function approach of \citet{hamba2004nonlocal} as discussed in \citet{liu2023systematic}.    

As an alternative to a computationally expensive brute force approach, inverse MFM can also be used to directly compute the moments of the eddy diffusivity in Equations \eqref{eq:D00}-\eqref{eq:D01} by forcing polynomial mean scalars \cite{mani2021macroscopic}. 
The eddy diffusivity moments need only one forced DNS per moment, and a few low-order moments are often sufficient
for quantification and modeling of nonlocal and anisotropic effects. Consider a simple one-dimensional (1D) example, in which averaging is taken over all directions except $x_1$ and there is only one component of the scalar flux, $\ens{u_1'c'}$. Equation \eqref{eq:kramers-moyal} becomes
\begin{align}
    \label{eq:Kramers-Moyal expansion moments}
    -\ens{u_1'c'}(x_1,t) = \left[D^{00}(x_1,t) + D^{10}(x_1,t)\frac{\partial}{\partial x_1} + D^{20}(x_1,t)\frac{\partial^2}{\partial x_1^2} + \dots + D^{01}(x_1,t)\frac{\partial}{\partial t} + \dots \right]\frac{\partial \ens{c}}{\partial x_1},
\end{align}
where we have omitted the \edit{tensor} subscripts on the moments, since here we only consider one spatial dimension.
To obtain the zeroth-order spatiotemporal moment of the eddy diffusivity, one specifies $\ens{c}=x_1$ using inverse MFM and solves the forced scalar transport equation in \eqref{eq:forced_ste}. 
At each time step, the forcing is used to maintain the specified $\ens{c}$ while $c$ is free to evolve. 
Practically, one can first time advance the governing equation without the forcing and solve for an intermediate scalar field, and then add the forcing in a correction step to ensure the scalar field at the next time step has the requisite $\ens{c}$ as discussed in \cite{mani2021macroscopic,bryngelson2024fast}. 
Postprocessing of $-\ens{u_1'c'}$ leads to the zeroth moment:
\begin{equation}
    -\ens{u_1'c'}|_{\ens{c}=x_1}(x_1,t) = D^{00}(x_1,t)
    \label{eq:1D example D00}
\end{equation} 
as shown by substitution of $\ens{c}=x_1$ into \eqref{eq:Kramers-Moyal expansion moments}. Specifying $\ens{c}$ as higher-order polynomials leads to higher-order moments of the eddy diffusivity. For the first-order spatial moment, specifying $\ens{c}=x_1^2/2$ gives
\begin{equation}
    -\ens{u_1'c'}|_{\ens{c}=x_1^2/2}(x_1,t) = x_1 D^{00}(x_1,t) + D^{10}(x_1,t)
    \label{eq:moment_calc_1}
\end{equation}
as shown by substitution of $\ens{c}=x_1^2/2$ into (\ref{eq:Kramers-Moyal expansion moments}). Post-processing the scalar flux and then subtracting out the contribution from the zeroth-order moment leads to $D^{10}$. Similarly, for the second-order spatial moment, specifying $\ens{c}=x_1^3/6$ gives
\begin{equation}
    -\ens{u_1'c'}|_{\ens{c}=x_1^3/6}(x_1,t) = \frac{x_1^2}{2} D^{00}(x_1,t) + x_1 D^{10}(x_1,t) + D^{20}(x_1,t),
\end{equation}
and post-processing the scalar flux and then subtracting out the contribution from the zeroth- and first-order spatial moments leads to $D^{20}$. 
For the first-order temporal moment, specifying $\ens{c}=x_1 t$ leads to
\begin{equation}
    -\ens{u_1'c'}|_{\ens{c}=x_1 t}(x_1,t) = tD^{00}(x_1,t) + D^{01}(x_1,t)
    \label{eq:1D example D01}
\end{equation} 
as shown by substitution of $\ens{c}=x_1 t$ into (\ref{eq:Kramers-Moyal expansion moments}).
Post-processing the scalar flux and then subtracting out the contribution from the zeroth-moment leads to $D^{01}$.

For a general multi-dimensional problem, other components of $D_{ij}^{mn}$ can be obtained by specifying the mean scalar in various coordinate directions and post-processing the components of the scalar flux. 
For example, specifying $\ens{c}=x_\alpha$ where $\alpha = 1, 2, \text{or } 3$ and substituting into the expansion in Equation \eqref{eq:kramers-moyal} gives
\begin{equation}
    -\ens{u_i'c'}(\mathbf{x},t)|_{\ens{c}=x_\alpha} = D_{i\alpha}^{00}(\mathbf{x},t).
\end{equation}
Postprocessing the scalar flux gives the $j=\alpha$ component of the zeroth-order spatiotemporal moment of the eddy diffusivity, $D_{i\alpha}^{00}$. Similarly, for the first-order spatial moment, specifying $\ens{c}=x_\alpha^2/2$ leads to 
\begin{equation}
    -\ens{u_i'c'}(\mathbf{x},t)|_{\ens{c}=x_\alpha^2/2} = x_\alpha D_{i\alpha}^{00}(\mathbf{x},t)+D_{i\alpha \alpha}^{10}(\mathbf{x},t),
    \label{eq:moment_calc_2}
\end{equation}
with no summation over $\alpha$ implied. Postprocessing the scalar flux and then subtracting out the contribution from the zeroth-order moment leads to $D_{i\alpha \alpha}^{10}$. Cross components of the first-order spatial moment may be obtained by specifying $\ens{c}=x_\alpha x_\beta$ where $\alpha, \beta = 1, 2, \text{or } 3$ and $\alpha \neq \beta$.

\begin{figure}
    \centering    
    \begin{subfigure}[]{\textwidth}
    \centering    
    \resizebox{.75\linewidth}{!}{
\begin{tikzpicture}[
node distance=2em, 
auto,
rect/.style={
    rectangle, 
    inner sep=1em
    },
rrect/.style={
    rect, 
    rounded corners=10
    },
smarrow/.style={
    arrows = {-Latex[width'=0pt .5, length=10pt, 
    line width=1.5pt]}
    }]
\tikzset{>=latex}
\pgfdeclarelayer{back}
\pgfsetlayers{back,main}

\node[rect] (base) {};
\node[below right=2em of base,rect,draw=seagreen,fill=white,line width=1.5pt, minimum width=12em] (n0) {$\frac{\partial c^{0}}{\partial t}=\mathcal{L}(c^{0})+s^{0}$};
\node[rect,fill=lightblue,draw=lightblue,above left=0.2em and 1.5em of n0] (donor1) {Donor};
\draw [arrows = {-Stealth[inset=0pt, angle=90:2em]},lightblue, line width=1em] (donor1) |- (n0)[midway];
\node[below=2em of n0,rect,draw=seagreen,fill=white,line width=1.5pt, minimum width=12em] (n1) {$\frac{\partial c^{1}}{\partial t}=\mathcal{L}(c^{1})+s^{1}$};
\node[rect,fill=lightblue,draw=lightblue,above left=0.2em and 1.5em of n1] (donor2) {Donor};
\draw [arrows = {-Stealth[inset=0pt, angle=90:2em]},lightblue, line width=1em] (donor2) |- (n1)[midway];
\node[below=2em of n1,rect,draw=seagreen,fill=white,line width=1.5pt, minimum width=12em] (n2) {$\frac{\partial c^{2}}{\partial t}=\mathcal{L}(c^{2})+s^{2}$};
\node[rect,fill=lightblue,draw=lightblue,above left=0.2em and 1.5em of n2] (donor3) {Donor};
\draw [arrows = {-Stealth[inset=0pt, angle=90:2em]},lightblue, line width=1em] (donor3) |- (n2)[midway];

\begin{pgfonlayer}{back}
\node[] at (base.north -| n0.north) (aligner) {};
\end{pgfonlayer};
\node[inner sep=0,above=1em of n0.north west,anchor=north west] (receivers) {\color{seagreen}Receiver};
\node[inner sep=0,above=1em of n1.north west,anchor=north west] (receivers) {\color{seagreen}Receiver};
\node[inner sep=0,above=1em of n2.north west,anchor=north west] (receivers) {\color{seagreen}Receiver};

\node[left=4.5em of n0.west] (s00) {$\langle c\rangle =x$};
\draw[smarrow] (s00) -> (n0);
\node[left=4.5em of n1.west] (s10) {$\langle c\rangle =\frac{x^2}{2}$};
\draw[smarrow] (s10) -> (n1);
\node[left=4.5em of n2.west] (s20) {$\langle c\rangle =\frac{x^3}{6}$};
\draw[smarrow] (s20) -> (n2);

\begin{pgfonlayer}{back}
\node[
    rrect,
    minimum height=20em,
    minimum width=7em,
    above right=1em and 19em of base.north,
    anchor=north,
    draw=black!15,
    line width=1.5pt,
    dashed
] (post_back) {};
\end{pgfonlayer};
\node[inner sep=0,below left=1em and 0em of post_back.north,anchor=north] (post) {\textcolor{black!50}{Postprocessing}};

\begin{pgfonlayer}{back}
\node[
    rrect,
    minimum height=20em,
    minimum width=19.5em,
    left=13.5em of post_back.north,
    anchor=north,
    draw=black!15,
    line width=1.5pt,
    dashed
] (dns_back) {};
\end{pgfonlayer};
\node[inner sep=0,below left=1em and 0em of dns_back.north,anchor=north] (dns) {\textcolor{black!50}{Multiple Simulations}};

\node[rect, draw=orange, line width=1.5pt, right=2.5em of n0.east] (F0) {$D^{0}$};
\draw[smarrow] (n0) -> (F0);
\node[rect, draw=orange, line width=1.5pt, right=2.5em of n1.east] (F1) {$D^{1}$};
\draw[smarrow] (n1) -> (F1);
\node[rect, draw=orange, line width=1.5pt, right=2.5em of n2.east] (F2) {$D^{2}$};
\draw[smarrow] (n2) -> (F2);

\end{tikzpicture}}
    \par\bigskip 
    \caption{MFM \edit{with separate donors}.}
    \label{fig:standard_MFM}
    \end{subfigure}
    \par\bigskip 
    \par\bigskip 
    \begin{subfigure}[]{\textwidth}
    \centering    
    \resizebox{.75\linewidth}{!}{
\begin{tikzpicture}[
node distance=2em, 
auto,
rect/.style={
    rectangle, 
    inner sep=1em
    },
rrect/.style={
    rect, 
    rounded corners=10
    },
smarrow/.style={
    arrows = {-Latex[width'=0pt .5, length=10pt, 
    line width=1.5pt]}
    }]
\tikzset{>=latex}
\pgfdeclarelayer{back}
\pgfsetlayers{back,main}

\node[rect] (base) {};
\node[below left=0em and 1.5em of base.north, rect,fill=lightblue,draw=lightblue,anchor=north] (donor) {Donor};
\node[below right=.75em and 1.5em of donor,rect,draw=seagreen,fill=white,line width=1.5pt, minimum width=12em] (n0) {$\frac{\partial \cmom^{0}}{\partial t}=\mathcal{L}^{0}(\cmom^{0})+s^{0}$};
\draw [arrows = {-Stealth[inset=0pt, angle=90:2em]},lightblue, line width=1em] (donor) |- (n0)[midway];
\node[below=2em of n0,rect,draw=seagreen,fill=white,line width=1.5pt, minimum width=12em] (n1) {$\frac{\partial \cmom^{1}}{\partial t}=\mathcal{L}^{1}(\cmom^{1})+s^{1}$};
\draw [arrows = {-Stealth[inset=0pt, angle=90:2em]},lightblue, line width=1em] (donor) |- (n1)[midway];
\node[below=2em of n1,rect,draw=seagreen,fill=white,line width=1.5pt, minimum width=12em] (n2) {$\frac{\partial \cmom^{2}}{\partial t}=\mathcal{L}^{2}(\cmom^{2})+s^{2}$};
\draw [arrows = {-Stealth[inset=0pt, angle=90:2em]},lightblue, line width=1em] (donor) |- (n2)[midway];

\begin{pgfonlayer}{back}
\node[] at (donor.north -| n0.north) (aligner) {};
\end{pgfonlayer};
\node[inner sep=0,above=1em of n0.north west,anchor=north west] (receivers) {\color{seagreen}Receiver};
\node[inner sep=0,above=1em of n1.north west,anchor=north west] (receivers) {\color{seagreen}Receiver};
\node[inner sep=0,above=1em of n2.north west,anchor=north west] (receivers) {\color{seagreen}Receiver};

\node[left=4.5em of n0.west] (s00) {$\frac{\partial \langle c\rangle}{\partial x}=1$};
\draw[smarrow] (s00) -> (n0);
\node[left=4.5em of n1.west] (s10) {$\frac{\partial \langle c\rangle}{\partial x}=x$};
\draw[smarrow] (s10) -> (n1);
\node[left=4.5em of n2.west] (s20) {$\frac{\partial \langle c\rangle}{\partial x}=\frac{x^2}{2}$};
\draw[smarrow] (s20) -> (n2);

\begin{pgfonlayer}{back}
\node[
    rrect,
    minimum height=20em,
    minimum width=7em,
    above right=1em and 19em of base.north,
    anchor=north,
    draw=black!15,
    line width=1.5pt,
    dashed
] (post_back) {};
\end{pgfonlayer};
\node[inner sep=0,below left=1em and 0em of post_back.north,anchor=north] (post) {\textcolor{black!50}{Postprocessing}};

\begin{pgfonlayer}{back}
\node[
    rrect,
    minimum height=20em,
    minimum width=19.5em,
    left=13.5em of post_back.north,
    anchor=north,
    draw=black!15,
    line width=1.5pt,
    dashed
] (dns_back) {};
\end{pgfonlayer};
\node[inner sep=0,below left=1em and 0em of dns_back.north,anchor=north] (dns) {\textcolor{black!50}{Single Simulation}};

\node[rect, draw=orange, line width=1.5pt, right=2.5em of n0.east] (F00) {$D^{0}$};
\draw[smarrow] (n0) -> (F00);
\node[rect, draw=orange, line width=1.5pt, right=2.5em of n1.east] (F10) {$D^{1}$};
\draw[smarrow] (n1) -> (F10);
\node[rect, draw=orange, line width=1.5pt, right=2.5em of n2.east] (F20) {$D^{2}$};
\draw[smarrow] (n2) -> (F20);

\end{tikzpicture}}
    \par\bigskip 
    \caption{\edit{MFM using a single donor and} decomposition method.}
    \label{fig:decomp_MFM}
    \end{subfigure}
    \caption{Diagrams outlining MFM \edit{implementation with separate donors vs.\ single donor} and the decomposition method in this work,  
    presented in one dimension for simplicity.
    Superscripts denote variables ($c^i$, $\cmom^i$, etc.) and operators ($\mathcal{L}^i$) belonging to the receiver equations solved to obtain $D^i$.}
    \label{fig:MFM_diagram}
\end{figure}
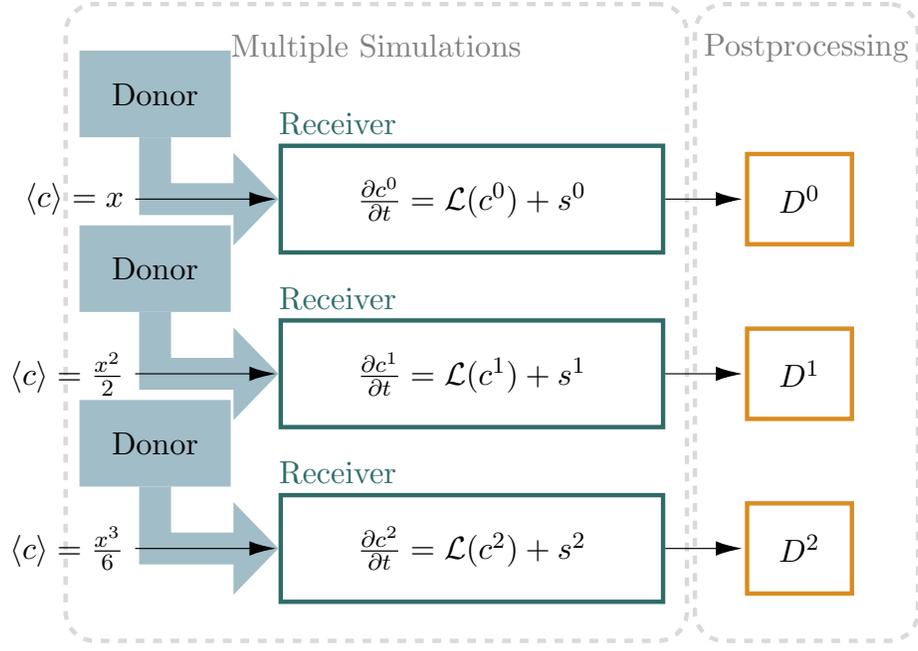
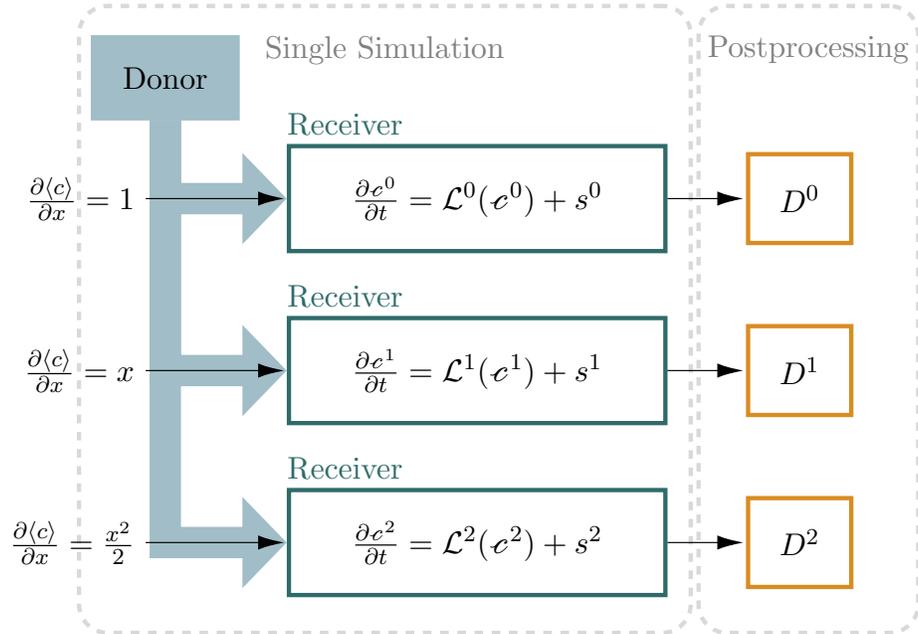

\begin{figure}
    \centering
    \begin{subfigure}[]{0.3\textwidth}
        \includegraphics[width=\textwidth]{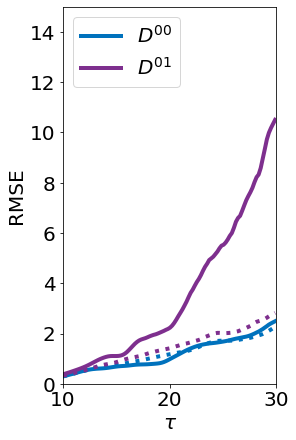}
        \subcaption[]{Leading order and first temporal moment.}
    \end{subfigure}
    \begin{subfigure}[]{0.3\textwidth}
        \includegraphics[width=\textwidth]{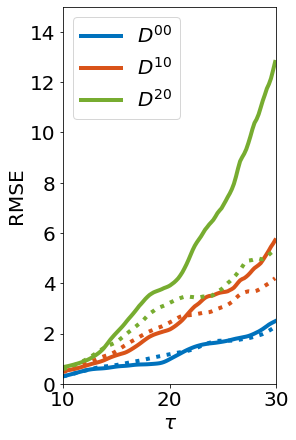}
        \subcaption[]{Leading order and spatial moments.}
    \end{subfigure}
    \caption{Root mean square errors (RMSE) of \edit{MFM with separate donors} (solid lines) vs.\ single donor (dotted lines) as percent of maximum magnitude of each eddy diffusivity moment at each time.
    Errors are computed at 20 realizations with respect to 200 realizations for each method.}
    \label{fig:stat_error}
\end{figure}

\subsection{Discussion of statistical error}
\label{sec:disc_stat_error}
A natural implementation of MFM for some codes is to use separate sets of donor equations \edit{(e.g., Navier--Stokes equations coupled with scalar transport equations)} and receiver equations \edit{(e.g., forced scalar transport equations (Equation \ref{eq:forced_ste}))} for each moment.
This configuration is illustrated in Figure \ref{fig:standard_MFM}.
This approach is more computationally expensive, since the donor equations are solved multiple times, but is easier to implement, since just one set of receiver equations is one-way coupled to each donor equation.

Another approach would be to run an MFM simulation using a single donor for multiple receivers, as illustrated in Figure \ref{fig:decomp_MFM} (the additional decomposition features of this proposed method will be detailed in a later section).
As we show in Appendix \ref{sec:error_analysis}, while mathematically equivalent to using a single donor, using separate donor simulations can result in larger statistical error.
Separate donor simulations, even when set up with identical initial conditions, may produce different instantaneous results due to  differences in the order of arithmetic operations (e.g., from parallelization), augmented by the chaotic nature of turbulence simulations.
\edit{As another example, for flows that require averaging over many realizations, differences may also arise in the donor simulations due to naive random generation of ensemble members.}
The resulting differences can propagate as error in the computation of higher-order eddy diffusivity moments from lower-order moments through Equations \eqref{eq:moment_calc_1}-\eqref{eq:moment_calc_2}, ultimately leading to slow statistical convergence.
When statistical errors between the donors match, the amplified error is removed, and statistical convergence is accelerated.
\editt{This can be ensured by using a single donor for multiple receivers. We note that in cases where the numerical solution between separate donors is identical, this approach is not necessary and using separate donors vs. a single donor will give the same result without amplification of error. However, using a single donor has the additional cost-saving advantage of not solving multiple instances of the donor equations and the added benefit of being able to use the decomposition method (detailed in Section \ref{sec:deco_form}) for problems where the MFM forcing may be incompatible with the boundary conditions of the problem. 
}

\editt{
To illustrate a case where separate donors can result in different numerical solutions due to differences in parallel processes amplified by the chaotic nature of turbulence,} Figure \ref{fig:stat_error} shows the statistical errors associated with \edit{using separate donors vs.\ a }single donor applied to the 2D RT problem \editt{in the \textit{Ares} code (detailed in Section \ref{sec:rt_sims})}.
In 2D RT instability, the flow is averaged over the homogeneous $x_1$ direction and over multiple realizations.
Details on the RT case study will be covered later in Section \ref{sec:RTI}, but we present these plots here to demonstrate the differences in error between the two methods.
When using \edit{Equation \eqref{eq:1D example D01}}, computation of $D^{01}$ involves multiplication of $D^{00}$ (and therefore its associated statistical error) by $t$, so we expect error to grow substantially at large $\tau$.
This is indeed what we observe in the statistical error plots.
Overall, across all higher-order moments, \edit{using separate donors} exhibits higher error than when using a single donor.
We show in Appendix \ref{sec:error_analysis} that when a single donor is used, errors from lower-order moments do not propagate to higher-order moments, so all errors scale similarly with time.

\section{Decomposition MFM}
\label{sec:deco_form}
In the previous section, we discussed the need for using the same DNS simulation (donor) for all forced MFM simulations (receivers) to reduce statistical error. In conjunction, we now introduce a decomposition method that can be used concurrently at no additional cost \edit{when the donor equations and all receiver equations are solved simultaneously}.
\edit{Mathematically equivalent to MFM, the} decomposition method treats the MFM forcing semi-analytically and was originally developed to address the issue of periodic boundary conditions in a steady laminar problem \cite{liu2023systematic}. Here we extend the decomposition to general unsteady and chaotic problems and to momentum transport in Section \ref{sec:momentum}.

Similar to the Green's function approach of \citet{hamba2004nonlocal}, we begin by substituting the Reynolds decomposition, $c(\mathbf{x}, t) = \ens{c}(\mathbf{x}, t) + c'(\mathbf{x}, t)$, into the scalar transport equation in \eqref{eq:ste} and subtracting the mean scalar transport equation to derive an equation for the scalar fluctuation, $c'$:
\begin{align}
    \frac{\partial c'}{\partial t} + \frac{\partial}{\partial x_i}\left(u_i c' - \ens{u_i'c'}\right) - D_M\frac{\partial^2c'}{\partial x_i \partial x_i} = -u_i'\frac{\partial \ens{c}}{\partial x_i}. 
    \label{eq:ste_fluc}
\end{align}
The general solution \cite{hamba2004nonlocal} for $c'(\mathbf{x},t)$ is 
\begin{align}
    c'(\mathbf{x},t) = \int\int g_{j}(\mathbf{x},\mathbf{x}',t,t') \left.\frac{\partial \ens{c}}{\partial x_j}\right|_{\mathbf{x}',t'} \text{d} \mathbf{x}' \text{d} t',
    \label{eq:gen_cp_soln}
\end{align}
where $g_{j}(\mathbf{x},\mathbf{x}',t,t')$ is the Green's function solution to
\begin{align}
    \frac{\partial g_j}{\partial t} + \frac{\partial}{\partial x_i}\left(u_i g_j - \ens{u_i'g_j}\right) - D_M\frac{\partial^2g_j}{\partial x_i \partial x_i} = -u_j'\delta(\mathbf{x} - \mathbf{x}')\delta(t - t').
\end{align}
As discussed in \citet{liu2023systematic}, the term $-\partial/\partial x_i \ens{u_i'c'}$ is related to the MFM forcing via
\begin{align}
    -\frac{\partial}{\partial x_i}\ens{u_i'c'} = \frac{\partial \ens{c}}{\partial t} + \ens{u_i}\frac{\partial \ens{c}}{\partial x_i} - D_M\frac{\partial^2\ens{c}}{\partial x_i\partial x_i} - s,
    \label{eq:forcing_relation}
\end{align}
and substitution of \eqref{eq:forcing_relation} into \eqref{eq:ste_fluc} leads to the forced transport equation for the scalar fluctuation:
\begin{align}
    \frac{\partial c'}{\partial t} + \frac{\partial}{\partial x_i}\left(u_i c'\right) - D_M\frac{\partial^2c'}{\partial x_i \partial x_i}  = s - \frac{\partial \ens{c}}{\partial t} - u_i\frac{\partial \ens{c}}{\partial x_i} + D_M\frac{\partial^2 \ens{c}}{\partial x_i \partial x_i}.
    \label{eq:forced_ste_fluc}
\end{align}
The purely mean scalar terms on the right hand side, $\partial \ens{c}/\partial t$ and $D_M \partial^2 \ens{c}/\partial x_i \partial x_i$, can be absorbed into the forcing, $s$, while still maintaining the property $s = \ens{s}$.
In this alternative formulation for inverse MFM, the forcing $s = \ens{s}$ is used to maintain $\ens{c'} = 0$. 
The right hand side term, $u_i \partial \ens{c}/\partial x_i$, is viewed as a source term to the scalar fluctuation equation. 
One advantage of using Equation \eqref{eq:forced_ste_fluc} rather than Equation \eqref{eq:forced_ste} for inverse MFM is that only the derivatives of the mean scalar, e.g., $\partial \ens{c}/
\partial t$ and $\partial \ens{c}/\partial x_i$, appear rather than the mean scalar, $\ens{c}$. The derivatives of the mean scalar are specified analytically and do not necessarily need a mathematically consistent mean scalar field. 
This point is further discussed in \citet{liu2024adjoint}, where we take advantage of the forced fluctuation equation to measure various components of the eddy viscosity tensor in turbulent channel flow independently.
The other advantage is that now we can further decompose $c'(\mathbf{x},t)$. 
Similar to the expansion in \eqref{eq:kramers-moyal}, consider the Taylor series expansion of the general solution for $c'(\mathbf{x},t)$ in \eqref{eq:gen_cp_soln} locally about $\mathbf{x}' = \mathbf{x}$ and $t = t'$:
\begin{align}
    c'(\textbf{x},t) = \left[\mathscr{c}_j^{00}(\mathbf{x},t) +
    \mathscr{c}_{jk}^{10}(\mathbf{x},t)\frac{\partial}{\partial x_k} + 
    \dots + 
    \mathscr{c}_j^{01}(\mathbf{x},t)\frac{\partial}{\partial t} +
    \dots\right] \frac{\partial \ens{c}}{\partial x_j},
    \label{eq:f expansion}
\end{align}
where 
\begin{align}
    \label{eq:f00}
    \mathscr{c}^{00}_j(\mathbf{x},t) &= \int \int g_j(\mathbf{x},\mathbf{x}',t,t')\text{d}\mathbf{x}' \text{d}t', 
    \\
    \label{eq:f10}
    \mathscr{c}^{10}_{jk}(\mathbf{x},t) &= \int \int (x_k'-x_k)g_j(\mathbf{x},\mathbf{x}',t,t')\text{d}x_k' \text{d}t',
    \\
    &\vdots \nonumber
    \\
    \label{eq:f01}
    \mathscr{c}^{01}_j(\mathbf{x},t) &= \int \int (t'-t)g_j(\mathbf{x},\mathbf{x}',t,t')\text{d}\mathbf{x}' \text{d}t',
    \\
    &\vdots \nonumber
\end{align}
\edit{
This expansion 
is similar to that used in moment-gradient methods for dispersion modeling, such as in the work of \citet{nadim1986higher}. 
In that work, the microscale probability density is expressed as an expansion of gradients of the macroscale probability density, analogous to our expansion of the microscopic $c'$ in terms of gradients of the macroscopic $\ens{c}$.
The moment-gradient expansion described by \citet{nadim1986higher} is used to derive dispersion models; in contrast, MFM does not directly output models but instead utilizes the expansion for numerical measurement of moments.
}

By substituting the decomposition \edit{in \eqref{eq:f expansion}} for $c'(\mathbf{x},t)$ into the forced scalar fluctuation equation in \eqref{eq:forced_ste_fluc}, we can derive governing equations for $\mathscr{c}^{mn}(\mathbf{x},t)$. As with inverse MFM, we can activate various $\mathscr{c}^{mn}(\mathbf{x},t)$ fields by specifying the mean scalar gradient. The main difference from inverse MFM as detailed in Section \ref{sec:MFM} is that the mean scalar gradient and MFM forcing are now treated semi-analytically. 
For example, specifying the mean scalar gradient as $\partial \ens{c}/\partial x_j = 1$ leads to an equation for $\mathscr{c}^{00}(\mathbf{x},t)$. Specifying the mean scalar gradient as higher-order polynomials leads to higher-order $\mathscr{c}^{mn}(\mathbf{x},t)$, and equations for the lower-order $\mathscr{c}^{mn}(\mathbf{x},t)$ can be analytically subtracted to derive an equation for the desired order of $\mathscr{c}^{mn}(\mathbf{x},t)$.
The $\mathscr{c}^{mn}(\mathbf{x},t)$ fields can then be used to compute the moments by multiplying \eqref{eq:f expansion} by $-u_i'$ and averaging:
\begin{align}
    -\ens{u_i'c'}(\textbf{x},t) = -  \left[\ens{u_i' \mathscr{c}_j^{00}}(\mathbf{x},t) +
    \ens{u_i' \mathscr{c}_{jk}^{10}}(\mathbf{x},t)\frac{\partial}{\partial x_k} + 
    \dots + 
    \ens{u_i' \mathscr{c}_j^{01}}(\mathbf{x},t)\frac{\partial}{\partial t} +
    \dots\right] \frac{\partial \ens{c}}{\partial x_j}.  
    \label{eq:f expansion moments}
\end{align}
Comparison with the expansion in \eqref{eq:kramers-moyal} leads to
\begin{align}
    -\ens{u_i'\mathscr{c}_j^{mn}}(\mathbf{x},t) = D_{ij}^{mn}(\mathbf{x},t). 
\end{align}

Consider the simple 1D example from Section \ref{sec:MFM}, where averaging is taken over all spatial directions except $x_1$ and there is only one component of the scalar flux, $\ens{u_1'c'}(x_1,t)$. To compute the zeroth-order moment, we specify $\partial \ens{c}/\partial x_1 = 1$ and substitution of the mean scalar gradient into \eqref{eq:f expansion} leads to:
\begin{align}
    c'(\mathbf{x},t) = \mathscr{c}^{00}(\mathbf{x},t),
    \label{eq:c_fluc_f00}
\end{align}
where we have omitted the subscript to only consider the $\mathscr{c} = \mathscr{c}_1$ component.
Substitution of \eqref{eq:c_fluc_f00} into the forced scalar fluctuation transport equation in \eqref{eq:forced_ste_fluc} leads to the equation for $\mathscr{c}^{00}(\mathbf{x},t)$:
\begin{align}
    \frac{\partial \mathscr{c}^{00}}{\partial t}+\frac{\partial}{\partial x_i}(u_i \mathscr{c}^{00})=D_M \frac{\partial^2  \mathscr{c}^{00}}{\partial x_i \partial x_i}-u_i\delta_{i1}+s^{00},
    \label{eq:f00 eqn}
\end{align}
where the inverse MFM forcing maintains the specified mean scalar gradient by equivalently maintaining $\ens{c'}(x_1,t) = \ens{\mathscr{c}^{00}}(x_1,t) = 0$. Postprocessing $-\ens{u_1'\mathscr{c}^{00}}(x_1,t)$ leads to the zeroth-order moment of the eddy diffusivity, $D^{00}(x_1,t)$.

To compute the first-order temporal moment, we specify $\partial \ens{c}/\partial x_1 = t$ and substitution of the mean scalar gradient into \eqref{eq:f expansion} leads to:
\begin{align}
    c'(\mathbf{x},t) = t\mathscr{c}^{00}(\mathbf{x},t) + \mathscr{c}^{01}(\mathbf{x},t),
    \label{eq:c_fluc_f01}
\end{align}
and substitution of the the decomposed scalar fluctuation in \eqref{eq:c_fluc_f01} into \eqref{eq:forced_ste_fluc} gives:
\begin{align}
    t\frac{\partial \mathscr{c}^{00}}{\partial t} + \mathscr{c}^{00} + \frac{\partial \mathscr{c}^{01}}{\partial t} + t\frac{\partial }{\partial x_i}(u_i \mathscr{c}^{00}) + \frac{\partial }{\partial x_i}(u_i \mathscr{c}^{01}) = D_M \left(t\frac{\partial^2  \mathscr{c}^{00}}{\partial x_i \partial x_i } + \frac{\partial^2  \mathscr{c}^{01}}{\partial x_i \partial x_i}\right) -tu_i\delta_{i1}+s.
    \label{eq:f01 eqn intermediate}
\end{align}
We analytically subtract the equation for $\mathscr{c}^{00}(\mathbf{x},t)$ in \eqref{eq:f00 eqn} multiplied by $t$ from Equation \eqref{eq:f01 eqn intermediate}
to arrive at an equation for $\mathscr{c}^{01}(\mathbf{x},t)$:
\begin{align}
    \frac{\partial \mathscr{c}^{01}}{\partial t}+u_i\frac{\partial \mathscr{c}^{01}}{\partial x_i}=D_M \frac{\partial^2  \mathscr{c}^{01}}{\partial x_i \partial x_i}-\mathscr{c}^{00}+s^{01},
\end{align}
where we have relabeled the inverse MFM forcing as $s^{01} = s - s^{00}t$. The forcing maintains $\ens{\mathscr{c}^{01}}(x_1,t) = 0$.
Postprocessing $-\ens{u_1'\mathscr{c}^{01}}(x_1,t)$ leads to the first-order temporal moment of the eddy diffusivity, $D^{01}(x_1,t)$.
Note the equation for $\mathscr{c}^{01}(\mathbf{x},t)$ is coupled with the equation for $\mathscr{c}^{00}(\mathbf{x},t)$. Generally, higher-order $\mathscr{c}^{mn}(\mathbf{x},t)$ are one-way coupled with lower-order $\mathscr{c}^{mn}(\mathbf{x},t)$.
\edit{Since higher-order $\mathscr{c}^{mn}(\mathbf{x},t)$ depend on lower-order $\mathscr{c}^{mn}(\mathbf{x},t)$, all decomposition receiver equations require access to the same donor solution, so the decomposition method naturally uses a single donor simulation.}

The cost of using the decomposition method is identical 
to the cost of MFM. By treating the forcing semi-analytically, the decomposition method can be used for periodic problems where the mean scalar gradient needed for MFM, e.g., $\partial \langle c \rangle/\partial x_1 = x_1$, may be incompatible with periodic boundary conditions. The equations for the decomposed variables satisfy the periodic boundary conditions and all explicit dependence on the coordinate is analytically removed. Moreover, the decomposition method also allows one to probe different directions of the eddy diffusivity independently. For example, consider a 2D problem, where $-\langle u_i'c' \rangle (x_1, x_2)$ and $\langle c \rangle(x_1, x_2)$. The decomposition method allows one to specify various directions of the mean scalar gradient, e.g., $\partial \langle c \rangle/\partial x_1 = x_2$ and $\partial \langle c \rangle/\partial x_2 = 0$, even when a $\langle c \rangle(x_1, x_2)$ that satisfies both desired gradients may not exist. This allows one to probe different directions of the eddy diffusivity, e.g., quantify $D_{11}$ independently from $D_{12}$. However, the resulting closure for $-\langle u_i'c \rangle$ is still mathematically consistent as a linear superposition of the various components of the eddy diffusivity.

\section{Generalization to momentum transport}
\label{sec:momentum}
The generalized closure can be extended to momentum transport, which is governed by the incompressible Navier--Stokes equations:
\begin{subequations}
    \begin{align}
        \label{eq:NS momentum}
        \frac{\partial u_i}{\partial t} + \frac{\partial}{\partial x_j}\left(u_ju_i\right) &= -\frac{1}{\rho}\frac{\partial p}{\partial x_i} + \nu \frac{\partial^2 u_i}{\partial x_j \partial x_j} + r_i, \\
        \label{eq:NS mass}
        \frac{\partial u_i}{\partial x_i} &= 0,
    \end{align}
\end{subequations}
where $p(\mathbf{x},t)$ is pressure, $\rho$ is the fluid density, $\nu$ is the kinetic viscosity, and $r_i(\mathbf{x},t)$ is a general body force.
Applying the Reynolds decomposition and averaging results in the Reynolds-averaged Navier--Stokes (RANS) equations:
\begin{align}
    \frac{\partial \ens{u_i}}{\partial t} + \frac{\partial}{\partial x_j}\left(\ens{u_j}\ens{u_i}\right) = -\frac{1}{\rho}\frac{\partial \ens{p}}{\partial x_i} + \nu \frac{\partial^2 \ens{u_i}}{\partial x_j \partial x_j} - \frac{\partial}{\partial x_j}\ens{u_j'u_i'} + \ens{r_i}.
\end{align}
The generalized nonlocal and anisotropic eddy viscosity~\cite{hamba2005nonlocal} is
\begin{align}
    \label{eq:generalized eddy viscosity}
    -\ens{u_i'u_j'}(\mathbf{x},t) = \int\int D_{ijkl}(\mathbf{x},\mathbf{x}',t,t') \left.\frac{\partial \ens{u_l}}{\partial x_k}\right|_{\mathbf{x}',t'} \mathrm{d}\mathbf{x}' \mathrm{d}t'.
\end{align}

To compute the generalized eddy viscosity, \citet{mani2021macroscopic} simultaneously solve the Navier--Stokes equations in \eqref{eq:NS momentum} and \eqref{eq:NS mass} and the generalized momentum (GMT) equations:
\begin{subequations}
    \begin{align}
        \label{eq:GMT momentum}
        \frac{\partial v_i}{\partial t} + \frac{\partial}{\partial x_j}\left(u_jv_i\right) &= -\frac{1}{\rho}\frac{\partial q}{\partial x_i} + \nu \frac{\partial^2 v_i}{\partial x_j \partial x_j} + s_i, \\
        \label{eq:GMT mass}
        \frac{\partial v_i}{\partial x_i} &= 0,
    \end{align}
\end{subequations}
where $v_i(\mathbf{x},t)$ is a vector field that is kept solenoidal by the scalar field $q(\mathbf{x},t)$, which acts similar to pressure, and $s_i$ is an added forcing that is not necessarily the same as $r_i$. The velocity field $u_j$ is computed from the Navier--Stokes equations, i.e., the GMT equations in \eqref{eq:GMT momentum} and \eqref{eq:GMT mass} are one-way coupled with the Navier--Stokes equations in \eqref{eq:NS momentum} and \eqref{eq:NS mass}.
The generalized closure \cite{hamba2005nonlocal} for the GMT equations is
\begin{align}
    \label{eq:generalized eddy viscosity v}
    -\ens{u_i'v_j'}(\mathbf{x},t) = \int\int D_{ijkl}(\mathbf{x},\mathbf{x}',t,t') \left.\frac{\partial \ens{v_l}}{\partial x_k}\right|_{\mathbf{x}',t'} \mathrm{d}\mathbf{x}' \mathrm{d}t'.
\end{align}

This closure is exact for the GMT equations, and the relationship between the closure operator in \eqref{eq:generalized eddy viscosity v} and in \eqref{eq:generalized eddy viscosity} is further discussed in \citet{mani2021macroscopic} and \citet{hamba2005nonlocal}.
\citet{park2024direct} numerically showed that substitution of the MFM-measured eddy viscosity kernel, $D_{ijkl} (\mathbf{x}, \mathbf{x}', t, t')$, from \eqref{eq:generalized eddy viscosity v} into \eqref{eq:generalized eddy viscosity} with the DNS mean velocity gradient results in  Reynolds stresses identical to DNS for turbulent channel flow.

Taking the Taylor series expansion of the nonlocal and anisotropic eddy viscosity in \eqref{eq:generalized eddy viscosity v} locally about $\mathbf{x}' = \mathbf{x}$ and $t' = t$:
\begin{equation}
    \label{eq:kramers-moyal channel}
    -\ens{u_{i}^{\prime} v_{j}^{\prime}}(\mathbf{x},t) = \left[ D^{00}_{ijkl}(\mathbf{x},t) + D^{10}_{ijklm}(\mathbf{x},t)\frac{\partial}{\partial x_m} + \dots + D^{01}_{ijkl}(\mathbf{x},t) \frac{\partial}{\partial t} + \cdots\right]\frac{\partial \ens{v_{l}}}{\partial x_{k}}
\end{equation}
where 
\begin{align}
    D_{ijkl}^{00} (\mathbf{x},t) &= \int \int D_{ijkl}(\mathbf{x},\mathbf{x}', t, t') \mathrm{d}\mathbf{x}'\mathrm{d}t', \\
    D_{ijklm}^{10} (\mathbf{x},t) &= \int \int (x_m' - x_m) D_{ijkl}(\mathbf{x}, \mathbf{x}', t,t')\mathrm{d}x_m' \mathrm{d}t', \\
    &\vdots \nonumber \\ 
    D_{ijkl}^{01} (\mathbf{x},t) &= \int \int (t' - t) D_{ijkl}(\mathbf{x}, \mathbf{x}', t,t')\mathrm{d}\mathbf{x}' \mathrm{d}t'.
\end{align}

Similar to scalar transport in Section \ref{sec:deco_form}, the MFM fluctuating velocity and pressure fields can be expanded as
\begin{equation}
    \label{eq:decomp v}
     v_{j}^{\prime}(\mathbf{x},t) = \left[ v^{\prime 00}_{jkl}(\mathbf{x},t) + v^{\prime 10}_{jklm}(\mathbf{x},t)\frac{\partial}{\partial x_m} + \dots + v^{\prime 01}_{jkl}(\mathbf{x},t) \frac{\partial}{\partial t} + \cdots\right]\frac{\partial \ens{v_{l}}}{\partial x_{k}},
\end{equation}
\begin{equation}
    \label{eq:decomp q}
     q(\mathbf{x},t) = \left[ q^{00}_{kl}(\mathbf{x},t) + q^{10}_{klm}(\mathbf{x},t)\frac{\partial}{\partial x_m} + \dots + q^{01}_{kl}(\mathbf{x},t) \frac{\partial}{\partial t} + \cdots\right]\frac{\partial \ens{v_{l}}}{\partial x_{k}}.
\end{equation}

For example, for a turbulent channel flow in which averaging is taken over the homogeneous streamwise ($x_1$) and spanwise ($x_3$) directions, the Reynolds stresses are only a function of the wall-normal ($x_2$) direction. The only nonzero component of the mean velocity gradient is $\partial\langle u_1 \rangle/\partial x_2$. To compute the zeroth moment of the generalized eddy viscosity using inverse MFM, one would specify $\partial \langle v_1 \rangle/\partial x_2 = 1$, and similar to scalar transport in Section \ref{sec:deco_form}, substitution of the specified mean velocity gradient into Equations \eqref{eq:GMT momentum}-\eqref{eq:GMT mass} and Equations \eqref{eq:decomp v}-\eqref{eq:decomp q}
leads to:
\begin{subequations}
    \begin{align}
        \label{eq:v00 momentum}
        \frac{\partial v_{j21}^{\prime 00}}{\partial t} + \frac{\partial}{\partial x_i}(u_i v_{j21}^{\prime 00}) &= -\frac{1}{\rho}\frac{\partial q_{21}^{00}}{\partial x_j} + \nu \frac{\partial^2  v_{j21}^{\prime 00}}{\partial x_i \partial x_i} - u_2 \delta_{j1} + s_j^{00}, \\
        \label{eq:v00 mass}
        \frac{\partial v_{j21}^{\prime 00}}{\partial x_j} &= 0,
    \end{align}
\end{subequations}
where inverse MFM is used to enforce $\langle v_{j21}^{\prime 00} \rangle=0$.
Postprocessing $-\langle u_i^\prime v_{j21}^{\prime 00} \rangle$ leads to the zeroth-order
moment of the eddy viscosity, $D_{ij21}^{00}$. 

For the first-order spatial moment in the wall-normal direction, substitution of $\partial \langle v_1 \rangle/\partial x_2 = x_2$ into Equations \eqref{eq:GMT momentum}-\eqref{eq:GMT mass} and \eqref{eq:decomp v}-\eqref{eq:decomp q} and subtraction of the equations for the zeroth-order moment in \eqref{eq:v00 momentum}-\eqref{eq:v00 mass} leads to
\begin{subequations}
    \begin{align}
        \label{eq:v10 momentum}
        \frac{\partial v_{j212}^{\prime 10}}{\partial t} + \frac{\partial}{\partial x_i}(u_i v_{j212}^{\prime 10}) &= -\frac{1}{\rho}\left[q_{21}^{00}\delta_{j2} + \frac{\partial q_{212}^{10}}{\partial x_j}\right] + \nu \left[2\frac{\partial v_{j21}^{\prime 00}}{\partial x_2} + \frac{\partial^2  v_{j212}^{\prime 10}}{\partial x_i \partial x_i}\right] - u_2 v_{j21}^{\prime 00} + s_j^{10}, \\
        \label{eq:v10 mass}
        \frac{\partial v_{j212}^{\prime 10}}{\partial x_j} &= -v_{221}^{\prime 00}.
    \end{align}
\end{subequations}
Inverse MFM is used to enforce $\langle v_{j212}^{\prime 10} \rangle=0$, and postprocessing $-\langle u_i^\prime v_{j212}^{\prime 10} \rangle$ leads to the first-order spatial
moment of the eddy viscosity, $D_{ij212}^{10}$. 
The continuity equation in \eqref{eq:v10 mass} is a direct result of substitution of the decomposition for $v_j'$ in Equation \eqref{eq:decomp v} into $\partial v_j'/\partial x_j = 0$. Generally, in cases where the specified mean velocity gradient is not solenoidal, we enforce the solenoidal condition on $v_j'$ rather than on $v_j$. As discussed in \citet{liu2024adjoint}, the specified mean velocity gradient can be considered as a MFM forcing to the continuity equation for $v_j$ that satisfies the requisite property $s = \langle s \rangle$. \edit{Appendix \ref{channel flow} shows computation of the zeroth-order and first-order spatial moments of the eddy viscosity for turbulent channel flow using decomposition MFM. Moreover, Appendix \ref{channel flow} shows that using the decomposition method has similar statistical errors as MFM with a single donor, which is expected since the two techniques are mathematically equivalent. In other words, Appendix \ref{channel flow} demonstrates that all improved statistical convergence is due to consolidation of donor simulations into a single donor.}

Similarly, for the first-order temporal moment, substitution of $\partial \langle v_1 \rangle/\partial x_2 = t$ into Equations \eqref{eq:GMT momentum}-\eqref{eq:GMT mass} and  \eqref{eq:decomp v}-\eqref{eq:decomp q} and subtraction of the equations for the zeroth-order moment in \eqref{eq:v00 momentum}-\eqref{eq:v00 mass} leads to
\begin{subequations}
    \begin{align}
        \frac{\partial v_{j21}^{\prime 01}}{\partial t} + \frac{\partial}{\partial x_i}(u_i v_{j21}^{\prime 01}) &= -\frac{1}{\rho}\frac{\partial q_{21}^{01}}{\partial x_j} + \nu \frac{\partial^2  v_{j21}^{\prime 01}}{\partial x_i \partial x_i} - v_{j21}^{\prime 00} + s_j^{01}, \\
        \frac{\partial v_{j21}^{\prime 01}}{\partial x_j} &= 0.
    \end{align}
\end{subequations}
Inverse MFM is used to enforce $\langle s_j^{01} \rangle=0$, and postprocessing $-\langle u_i^\prime v_{j21}^{\prime 01} \rangle$ leads to the first-order temporal
moment of the eddy viscosity, $D_{ij21}^{01}$. 
\citet{park2022snh} computed the first-order temporal moment for turbulent channel flow at $Re_\tau=180$. They then used the temporal moment as a qualitative estimate for a nonlocality timescale in a simple nonlocal model for a 2D separated boundary layer.  

As was the case for scalar transport, the equations for momentum transport for higher-order spatiotemporal moments generally depend on lower-order moments, which are solved simultaneously. Equations for higher-order moments are one-way coupled with lower-order moments and do not raise the cost of the MFM procedure. 

\section{Case study: Rayleigh--Taylor instability}
\label{sec:RTI}
As an illustrative case study, we demonstrate the decomposition method for mean scalar transport in two-dimensional (2D) RT instability. 
RT instability occurs when a heavier fluid is accelerated into a lighter fluid with a perturbation at the interface of the two fluids.
Over time, the instability becomes self-similar and enters a turbulent state.
RT instability is a chaotic, non-stationary flow, so statistical convergence must be achieved through ensemble averaging.
In the 2D RT problem, the only homogeneous direction is $x_1$, and there is no homogeneity in time that can be leveraged for ensemble averaging.
Thus, many realizations of RT instability, each with different initial conditions, are required to get statistical convergence of the eddy diffusivity moments.
This corresponds to many DNSs, which lends to the high computational expense of MFM for this problem.

Past work \citep{lavacot2024} showed that, using standard MFM \edit{with separate donors}, $\mathcal{O}(10^3)$ DNSs are required for statistical convergence of eddy diffusivity moments in 2D RT instability.
In this case study, the decomposition MFM is applied to \edit{the same }2D RT instability \edit{problem}.
\edit{Since decomposition MFM naturally requires a single donor equation, we} achieve faster statistical convergence of the eddy diffusivity moments \edit{than with MFM using separate donors}.
\edit{
We emphasize that the improved statistical convergence is wholly due to the use of a single donor, not the decomposition.
For a case study investigating the separate effects of using a single donor and using the decomposition on statistical convergence, we refer the reader to the turbulent channel flow study in Appendix \ref{channel flow}. 
}

\edit{
For this study, we specifically consider scalar transport in compressible RT instability.
Note that this means that the governing equations are not Equations \eqref{eq:NS momentum} and \eqref{eq:NS mass}, but are instead described later in Section \ref{sec:rt_goveqs}.
The relevant decomposition formulation for the scalar transport problem is described in Section \ref{sec:deco_form}.
}

\subsection{Governing equations}
\label{sec:rt_goveqs}
The compressible Navier--Stokes equations are solved in the donor simulation:

\begin{align}
    \frac{D\rho}{Dt}&=-\rho\frac{\partial u_i}{\partial x_i}\label{eq:rti_donor1},\\
    \rho\frac{DY_k}{Dt}&=\frac{\partial }{\partial x_i}\left(\rho D_k\frac{\partial Y_k}{\partial x_i}\right),\label{eq:rti_scalar_transp}\\
    \rho\frac{Du_j}{Dt}&=-\frac{\partial }{\partial x_i}\left(p\delta_{ij}+\sigma_{ij}\right) + \rho g_j,\\
    \rho\frac{De}{Dt}&=-p\frac{\partial u_i}{\partial x_i}+\frac{\partial }{\partial x_j}\left(u_i\sigma_{ij}-q_j\right)\label{eq:rti_donor2},
\end{align}
where $\rho$ is density, $u_i$ is velocity, $Y_k$ is mass fraction of component $k$, $D_k$ is the molecular diffusivity of component $k$ (in the problem considered here, $D_k=D_M$), $p$ is pressure, $g_j$ is gravity (only the $x_2$ component is active in this problem), and $e$ is specific internal energy.
The viscous stress $\sigma_{ij}$ and the heat flux $q_{ij}$ are
\begin{align}
    \sigma_{ij} &= \mu\left(\frac{\partial u_i}{\partial x_j}+\frac{\partial u_j}{\partial x_i}\right)-\mu\frac{2}{3}\frac{\partial u_m}{\partial x_m}\delta_{ij},\\
    q_j &= -\kappa\frac{\partial T}{\partial x_j} - \sum^N_{k=1}h_k\rho D_k\frac{\partial Y_k}{\partial x_j},
\end{align}
where $\mu$ is the dynamic viscosity, $\kappa$ is the thermal conductivity, $T$ is temperature, and $h_k$ is the specific enthalpy of species $k$.
Component pressures are determined using ideal gas equations of state.
The total pressure is a weighted sum of component pressures:
\begin{align}
    p&=\sum^N_{k=1}v_k p_k,
\end{align}
\edit{where $v_k$ is the species volume fraction.}
More details on these equations can be found in \citet{lavacot2024} and \citet{morgan2018large}.

\subsection{Self-similarity}
After transition to turbulence, the growth of the RT mixing layer becomes self-similar in time.
In this limit, $h$, defined as half of the width of the mixing layer, is expected to grow quadratically in time:
\begin{align}
    h=\alpha A g t^2,
\end{align}
where $\alpha$ is the growth parameter and $A$ is the Atwood number, defined as
\begin{align}
    A = \frac{\rho_H-\rho_L}{\rho_H+\rho_L},
    \label{eq:Atwood}
\end{align}
where $\rho_H$ is the density of the heavy fluid, and $\rho_L$ is the density of the light fluid.
In this work, we perform the MFM analysis in this self-similar limit.
We define the self-similar variable:
\begin{align}
    \eta = \frac{x_2}{h(t)}.
\end{align}

The mixing width can be computed from \edit{$Y_H$}, the mass fraction \edit{of the heavy fluid}:
\begin{align}
    h \equiv 4\int\langle Y_H\left(1-Y_H\right) \rangle dx_2,
\label{eq:h_meas}
\end{align}
where $\langle*\rangle$ denotes an ensemble average; for this 2D RT problem, the averaging is done in $x_1$ and over realizations.
In a RANS simulation, the mixing width can instead be redefined using closed quantities, as defined by \citet{cabot2006reynolds} and \citet{morgan2017self}:
\begin{align}
    h_\text{hom} \equiv 4\int\langle Y_H \rangle \left(1-\langle Y_H \rangle \right)dx_2.
    \label{eq:h_hom}
\end{align}
\edit{
Additionally, the mixing half-width can also be computed from mass fraction profiles by taking it as the distance from the centerline of the domain to where the mean mass fraction of the light fluid is $0.999$.
This definition of the mixing half-width is referred to as $h_{99}$.
}

A mixedness parameter $\phi$ can also be defined (\citet{youngs1994numerical, morgan2017self}):
\begin{align}
    \phi \equiv \frac{h}{h_\text{hom}}=1-4\frac{\int\langle Y_H'Y_H' \rangle dx_2}{h_\text{hom}}.
\end{align}
The mixedness $\phi$ is expected to converge to a steady-state value in the self-similar limit.

\subsection{Computation of eddy diffusivity moments}
For RT instability, after averaging over the homogeneous $x_1$ direction, the only surviving turbulent flux is $-\ens{u_2'c'}$, where $c$ is the mass fraction of the heavy fluid, $Y_H$.
The Kramers--Moyal expansion in Equation \eqref{eq:kramers-moyal} becomes
\begin{align}
    -\ens{u_2'c'}(\textbf{x},t) = D^{00}\frac{\partial \ens{c}}{\partial x_2} +
    D^{10}\frac{\partial^2 \ens{c}}{\partial x_2^2} +
    D^{01}\frac{\partial^2 \ens{c}}{\partial t \partial x_2} +
    D^{20}\frac{\partial^3 \ens{c}}{\partial x_2^3} + \dots 
    \label{eq:RTI_moments}
\end{align}
In this work, $D^{00}$, $D^{01}$, $D^{10}$, and $D^{20}$ are computed.

\begin{table}[]
    \centering
    {\renewcommand{\arraystretch}{1.15}%
    \begin{tabular}{c l}
         Moment &  $\frac{\partial Y_H}{\partial x_2}$\\
         \hline
         $D^{00}$ & $1$\\
         $D^{01}$ & $t$\\
         $D^{10}$ & $x_2$\\
         $D^{20}$ & $\frac{1}{2}x_2^2$\\ 
    \end{tabular}}
    \caption{Mean mass fraction gradients forced for each eddy diffusivity moment $D^{mn}$ in the 2D RT case study\edit{, where $x_2$ is defined between $-\frac{1}{2}$ and $\frac{1}{2}$}.}
    \label{tab:moment_forcings}
\end{table}

In standard MFM, the macroscopic forcing would be applied directly to Equation \eqref{eq:rti_scalar_transp}, and the eddy diffusivity moments would be obtained in postprocessing.
For example, to compute $D^{10}$, a macroscopic forcing to Equation \eqref{eq:rti_scalar_transp} would be determined to enforce \edit{$ \ens{Y_H} = x_2^2/2$.}
From the solution to that receiver equation, the moment is computed as $D^{10}=-\ens{u_2'c'}|_{\ens{Y_H}= x_2^2/2} - x_2D^{00}$.
The forcings for the other moments are shown in Table \ref{tab:moment_forcings}.

In this study, we compare the statistical convergence of eddy diffusivity moments computed using standard MFM \edit{with separate donors} and decomposition MFM.
According to the decomposition described in Section \ref{sec:deco_form}, we derive four receiver equations for the decomposition MFM:

\begin{align}
& \frac{\partial \cmom^{00}}{\partial t}+u_i\frac{\partial \cmom^{00}}{\partial x_i}=D_M \frac{\partial^2  \cmom^{00}}{\partial x_i \partial x_i}-u_i\delta_{i2}+s^{00},
\label{eq:rti_receiver1} \\
& \frac{\partial \cmom^{10}}{\partial t}+u_i\frac{\partial \cmom^{10}}{\partial x_i}=D_M \frac{\partial^2  \cmom^{10}}{\partial x_i \partial x_i}+D_M\left(1+2 \frac{\partial \cmom^{00}}{\partial x_i}\delta_{i2}\right)-u_i\delta_{i2} \cmom^{00}+s^{10}, \\
& \frac{\partial \cmom^{01}}{\partial t}+u_i\frac{\partial \cmom^{01}}{\partial x_i}=D_M \frac{\partial^2  \cmom^{01}}{\partial x_i \partial x_i}-\cmom^{00}+s^{01}, \\
& \frac{\partial \cmom^{20}}{\partial t}+u_i\frac{\partial \cmom^{20}}{\partial x_i}=D_M \frac{\partial^2  \cmom^{20}}{\partial x_i \partial x_i}+D_M\left(\cmom^{00}+2 \frac{\partial \cmom^{10}}{\partial x_i}\delta_{i2}\right)-u_i\delta_{i2} \cmom^{10}+s^{20}. 
\label{eq:rti_receiver2}
\end{align}
The forcings $s^{ij}$ are determined to enforce zero means in $x_1$ for each $\cmom^{ij}$ in each realization.
These forcings are computed per timestep in each realization, as detailed in \citet{lavacot2024}.
With this formulation, the forcings in Table \ref{tab:moment_forcings} are now semi-analytically applied.
Each moment is computed in postprocessing:
\begin{align}
    D^{mn}=-\ens{ u_2'\cmom^{mn} }.
\end{align}

\subsection{Simulations}
\label{sec:rt_sims}
The hydrodynamics solver Ares \cite{morgan2016large, bender2021simulation} is used to run 2D RT simulations.
Ares uses an arbitrary Lagrangian-Eulerian (ALE) method based on \citet{sharp1981hemp}.
In this method, equations are solved in a Lagrangian frame and then remapped to an Eulerian mesh using a second-order scheme.
Ares uses a second-order non-dissipative finite element method in space and a second-order explicit predictor-corrector scheme in time.

MFM is performed two different ways to measure eddy diffusivity moments for the 2D RT instability.
The first is MFM \edit{using separate donors}.
In this case, for each realization, four receiver equations are solved alongside four separate donor equations, but all use the same initial conditions.
The second case \edit{uses the decomposition MFM, which requires a single donor}, as described in Section \ref{sec:deco_form}.
In the decomposition MFM, for each realization of RT instability, the four receiver equations (Equations \eqref{eq:rti_receiver1}-\eqref{eq:rti_receiver2}) are solved alongside one set of donor equations (Equations \eqref{eq:rti_donor1}-\eqref{eq:rti_donor2}).

The 2D simulations are run on a square domain of $2049\times2049$ cells with periodic boundary conditions in $x_1$ and no slip and no penetration \edit{at $x_2=-\frac{1}{2}$ and $\frac{1}{2}$}.
To trigger the instability, a top-hat perturbation in wavespace between the heavy and light fluids is applied to the density field \edit{at $x_2=0$}.
The perturbation has a minimum wavenumber $\kappa_{min}=8$, a maximum wavenumber $\kappa_{max}=256$, and an amplitude of $\frac{\Delta}{\kappa_{max}-\kappa_{min}+1}$, where $\Delta$ is the grid size.
The simulations are stopped when the mixing width is approximately $30\%$ the size of the domain.

{
\renewcommand{\arraystretch}{1.5}
\begin{table}[]
    \centering
    \begin{tabular}{c|c|c}
         Number & Definition & Value \\
         \hline
         $A$ & Equation $\eqref{eq:Atwood}$&  $0.05$\\
         ${Ma}_\text{max}$ & $\frac{\sqrt{u_iu_i}}{c}$&  $0.05$\\
         $Gr$ & $\frac{-2gA\Delta^3}{\nu^2}$&  $1$\\
         $Sc$ & $\frac{\nu}{D_M}$&  $1$\\ 
         $Re_T$ & $\frac{k^{1/2}\lambda}{\nu}$&  $54$\\
         $Pe_T$ & $Re_T Sc$&  $54$\\
         $Re_L$ & $\frac{h_{99}\frac{d{h}_{99}}{dt}}{\nu}$&  $8,000$\\
         $Pe_L$ & $ Re_L Sc$&  $8,000$\\
    \end{tabular}
    \caption{Nondimensional numbers of simulated RT instability \edit{at the last timestep of the simulations}. Here, $c$ is the speed of sound (set by the heat capacity ratio $\gamma$, which is $5/3$ in the simulation), $\Delta$ is the grid spacing (the mesh is uniform, so $\Delta=\Delta_x=\Delta_y$), and $D_M$ is the molecular diffusivity.
    The subscripts $T$ and $L$ refer to nondimensional numbers using the Taylor microscale ($\lambda$) and large-scale, respectively.
    \edit{
    The turbulent kinetic energy $k$ is defined as $\frac{1}{2}\ens{u_i'u_i'}$.
    }
    }
    \label{tab:RTI_nondim}
\end{table}
}

The relevant nondimensional numbers of this problem are the Atwood number ($A$), the Reynolds number ($Re$) (which is set in the simulation by a numerical Grashof number ($Gr$)), Mach number ($Ma$), Peclet number ($Pe$), and Schmidt number ($Sc$).
These are defined in Table \ref{tab:RTI_nondim}.
The RT flow can be considered turbulent when $Re_T>100$ or $Re_L>10,000$ \citep{dimotakis2000mixing}.
Details on how these numbers are computed for the donor simulation can be found in \citet{lavacot2024}.

Since the Atwood number is small, the bouyant-flow Boussinesq approximation can be made.
Thus, it is assumed that mean velocities are negligible.
In addition, a small Grashof number based on mesh size is used to minimize numerical diffusion and keep the simulation close to a DNS.
It was found by \citet{morgan2020parametric} that numerical diffusivity dominates molecular diffusivity when $Gr>12$.
Finally, it must be noted that $Re_T$ and $Re_L$ of the RT instability simulated here are lower than the turbulent transition criteria set by \citet{dimotakis2000mixing}.
This indicates that the flow may not be fully turbulent; in fact, the simulation is 2D, so it cannot be truly turbulent.
However, the late-time profiles from the simulations show self-similar behavior (see \citet{lavacot2024}), so self-similar analysis is valid.

\subsection{Donor solution}

\begin{figure}
    \centering
    \includegraphics[width=0.5\linewidth]{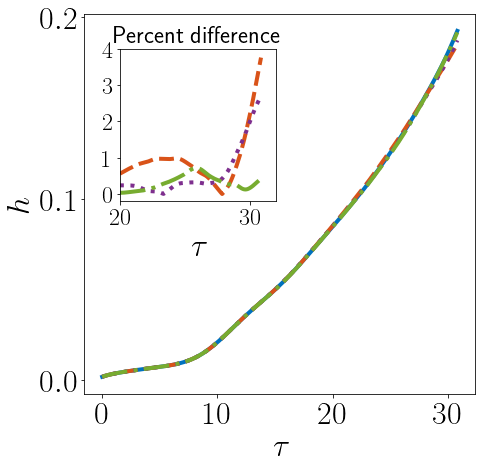}
    \caption{Mixing half-width $h$ measured from \edit{separate} donor simulations in Ares for the same initial conditions. 
    Inset plot is the percent differences of the last three donors with respect to the first donor.
    $\tau$ is a nondimensional time, defined as $\frac{t}{t_0}$, where $t_0=\sqrt{\frac{h_0}{Ag}}$ and $h_0$ is the dominant lengthscale determined by the peak of the initial perturbation spectrum.}
    \label{fig:diff_donors}
\end{figure}

Figure \ref{fig:diff_donors} shows the mixing half-widths measured from the standard MFM donor simulations over time.
In applying the standard MFM to compute $D^{00}$, $D^{10}$, $D^{01}$, and $D^{20}$, four \edit{separate} donor simulations are used.
These donors use the same initial conditions, but due to numerical differences in parallel processes, Ares gives slightly different $h$ at late time.
For $h$, the percent error is only $\mathcal{O}(1\%)$, but the statistical error is amplified in the computation of higher-order eddy diffusivity moments, as discussed previously in Section \ref{sec:disc_stat_error}.
A more detailed description of the donor simulations can be found in \citet{lavacot2024}.

\subsection{Eddy diffusivity moments}
The solutions to the receiver equations are postprocessed to obtain the eddy diffusivity moments.
\begin{figure}
    \centering
    \begin{subfigure}[]{0.49\textwidth}
        \includegraphics[width=\textwidth]{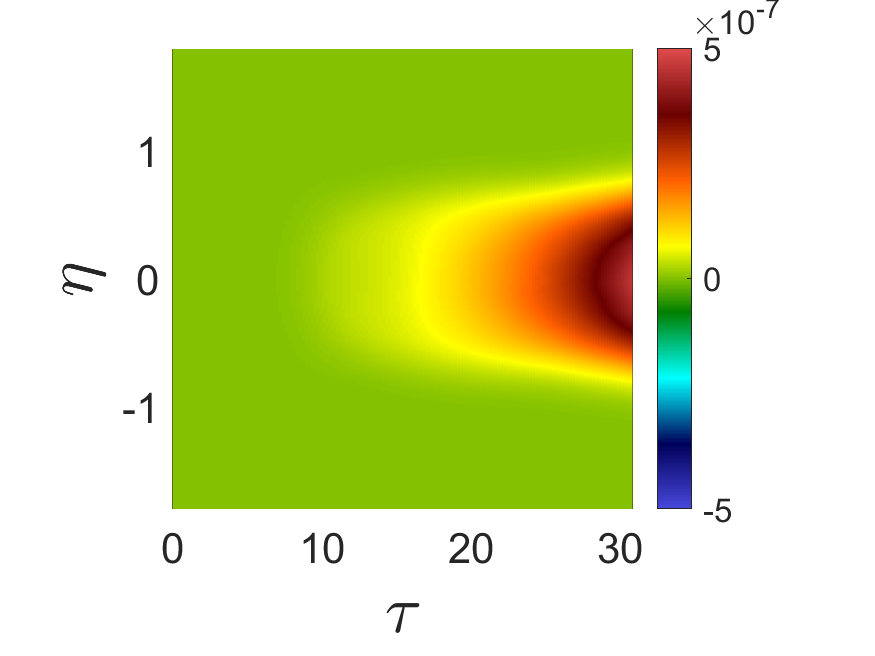}
        \subcaption[]{$D^{00}$}
    \end{subfigure}
    \begin{subfigure}[]{0.49\textwidth}
        \includegraphics[width=\textwidth]{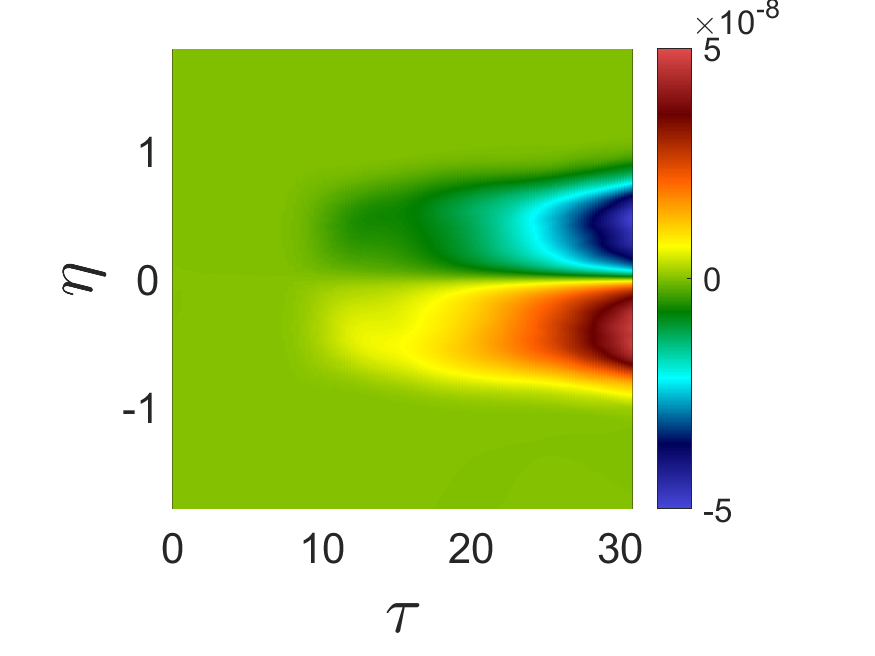}
        \subcaption[]{$D^{10}/h(t)$}
    \end{subfigure}
    \begin{subfigure}[]{0.49\textwidth}
        \includegraphics[width=\textwidth]{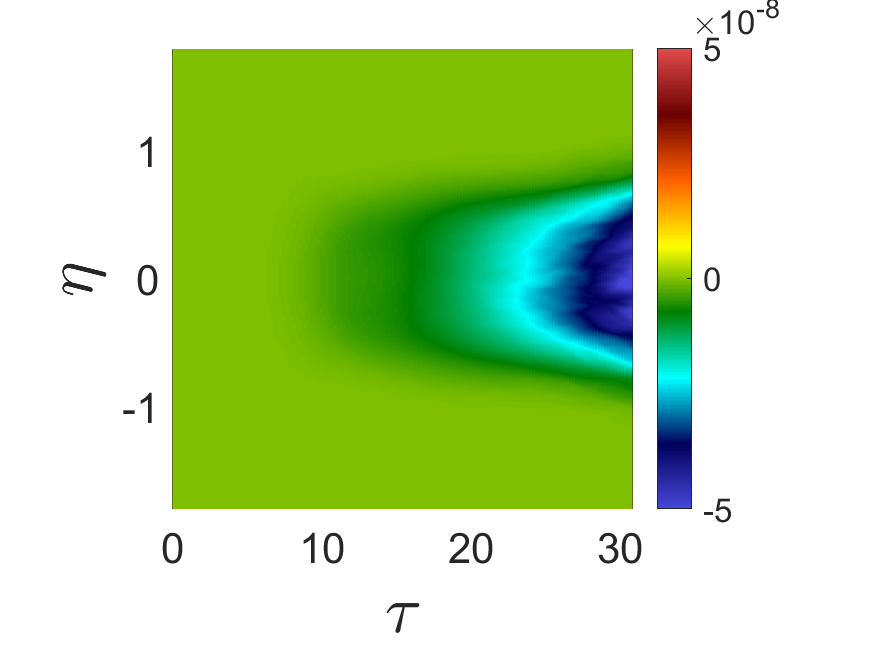}
        \subcaption[]{$D^{01}/t$}
    \end{subfigure}
    \begin{subfigure}[]{0.49\textwidth}
        \includegraphics[width=\textwidth]{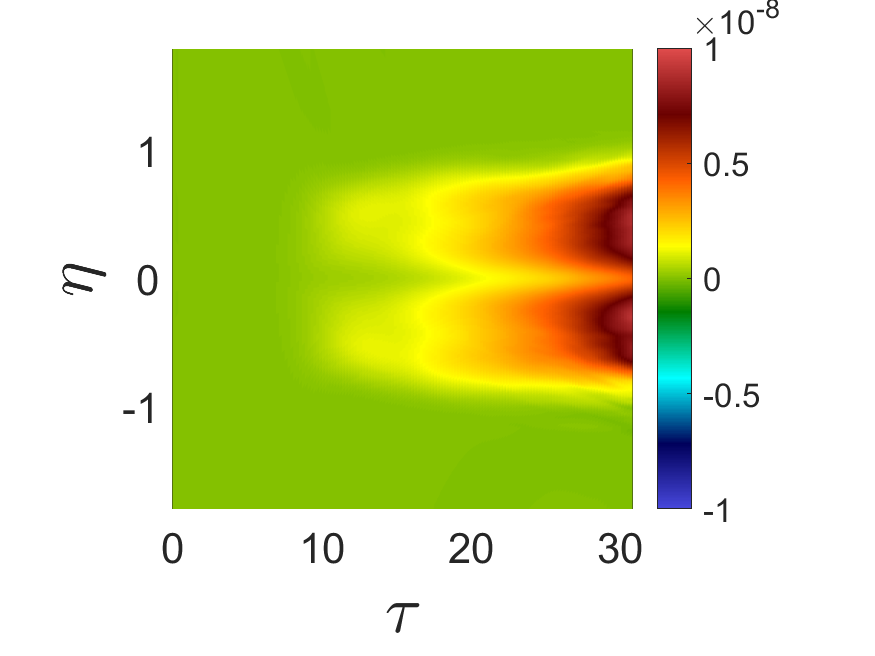}
        \subcaption[]{$D^{20}/h(t)^2$}
    \end{subfigure}
    \caption{Normalized moments of the eddy diffusivity kernel of RT instability measured using the standard MFM \edit{with separate donors}. Data is averaged over 1,000 realizations and homogeneous direction $x_1$.}
    \label{fig:RTI_moments_orig}
\end{figure}
\begin{figure}
    \centering
    \begin{subfigure}[]{0.49\textwidth}
        \includegraphics[width=\textwidth]{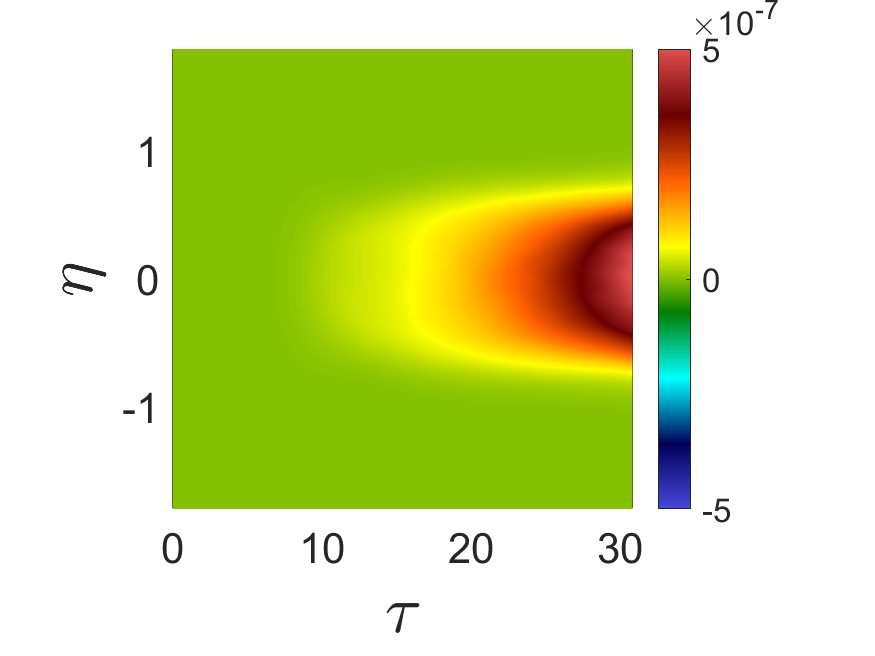}
        \subcaption[]{$D^{00}$}
    \end{subfigure}
    \begin{subfigure}[]{0.49\textwidth}
        \includegraphics[width=\textwidth]{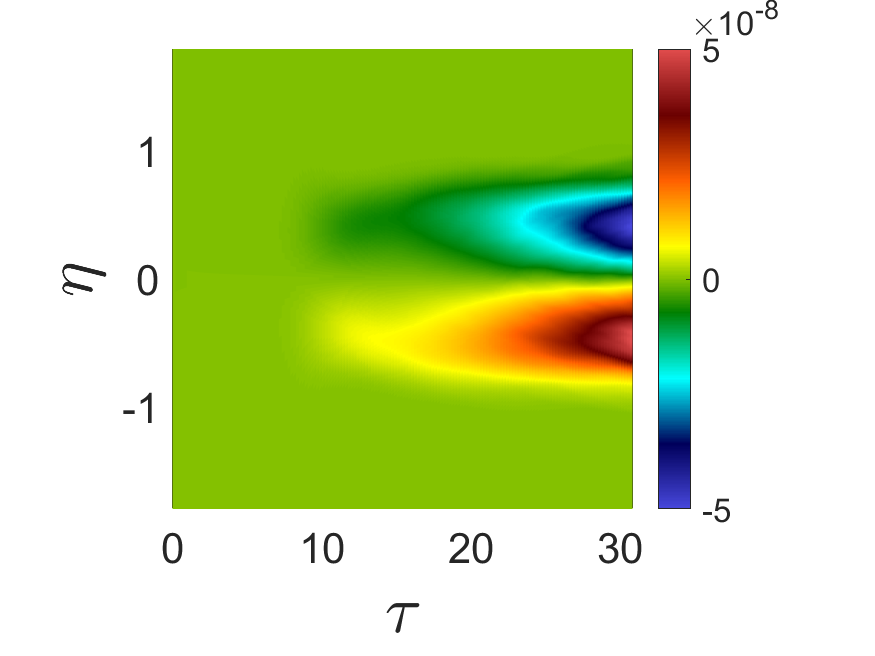}
        \subcaption[]{$D^{10}/h(t)$}
    \end{subfigure}
    \begin{subfigure}[]{0.49\textwidth}
        \includegraphics[width=\textwidth]{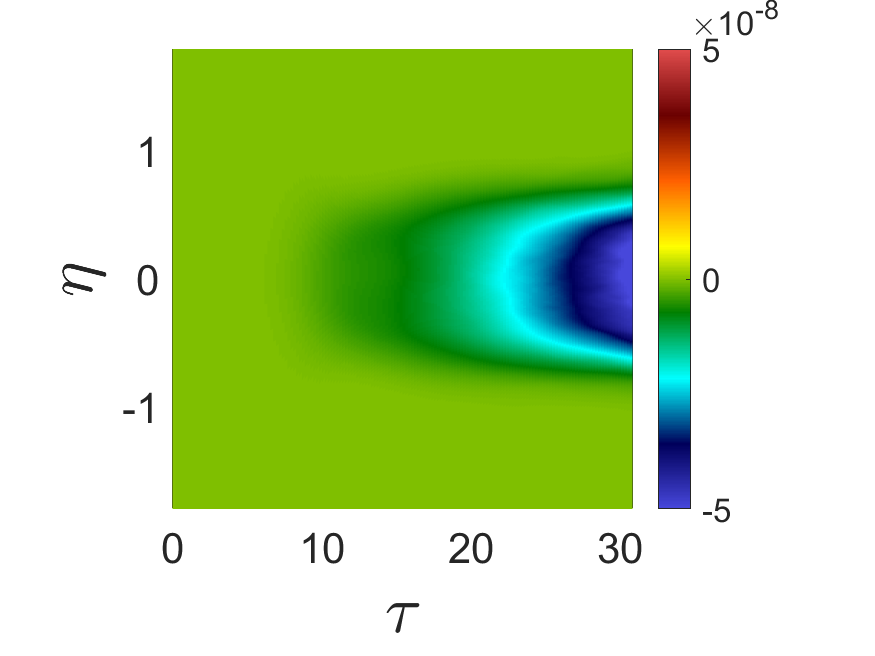}
        \subcaption[]{$D^{01}/t$}
    \end{subfigure}
    \begin{subfigure}[]{0.49\textwidth}
        \includegraphics[width=\textwidth]{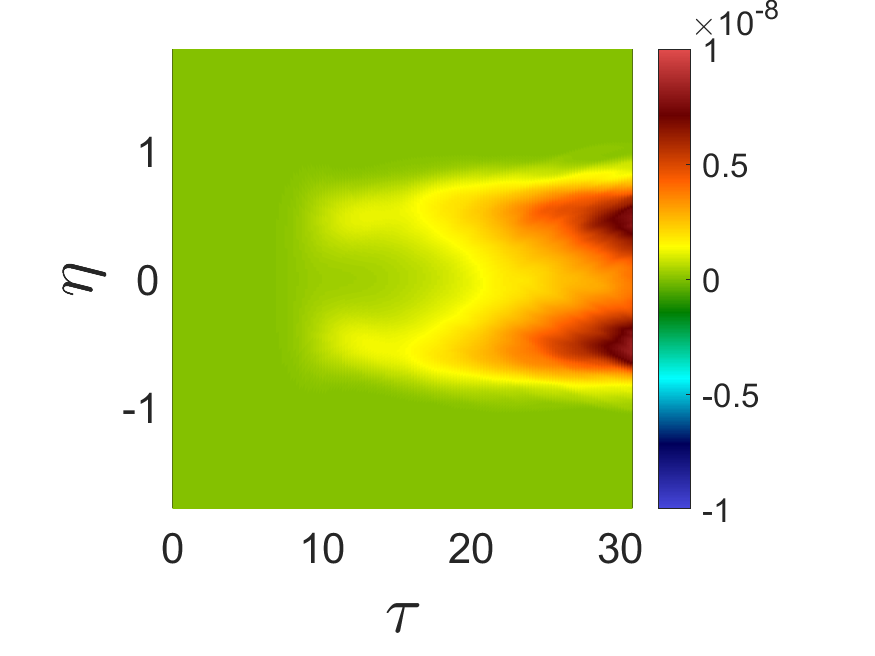}
        \subcaption[]{$D^{20}/h(t)^2$}
    \end{subfigure}
    \caption{Normalized moments of the eddy diffusivity kernel of RT instability measured using the decomposition MFM. Data is averaged over 200 realizations and homogeneous direction $x_1$.}
    \label{fig:RTI_moments_decomp}
\end{figure}
Figure \ref{fig:RTI_moments_orig} shows eddy diffusivity moments measured using standard MFM \edit{with separate donors}.
The data are averaged over $1,000$ realizations.
Even with this large number of realizations, the measurements exhibit substantial statistical error, especially in the higher-order moments; $D^{01}$, the first order moment in time is particularly affected by this issue.
Figure \ref{fig:RTI_moments_decomp} shows eddy diffusivity moments measured using the decomposition MFM\edit{, which uses a single donor}. 
The $D^{00}$ measurements using the two methods are qualitatively similar and have about the same level of statistical error.
This is expected, since the calculations for the leading order moment in either method are mathematically equivalent.
Among the higher-order moments, the decomposition MFM measurements show significantly improved statistical convergence at only $200$ averaged realizations.
This improvement is most noticeable in the measurement of $D^{01}$.
Compared to the measurement \edit{from standard MFM using separate donors}, the decomposition MFM measurement is smoother and more symmetric, qualitatively indicating less statistical error.

\begin{figure}
    \centering
    \resizebox{.8\linewidth}{!}{\input{RTI_convergence.tex}}
    \caption{Convergence of $D^{01}$ (normalized by $t$)  measurement using standard \edit{MFM with separate donors} and decomposition \edit{MFM with a single donor} for RT instability case.}
    \label{fig:RTI_convergence}
\end{figure}

Since the statistical error is most obvious in the measurements of $D^{01}$, those measurements averaged over different numbers of realizations are presented in Figure \ref{fig:RTI_convergence}.
Even at just one realization, the decomposition MFM measurement exhibits much less statistical error than the standard MFM \edit{using separate donors}.
As the number of realizations increases, the statistical error reduces much faster in the decomposition MFM at $100$ realizations.
Visually, the decomposition MFM measurement has an acceptable level of statistical error at only $100$ averaged realizations, but the standard MFM \edit{with separate donors} still has a high level of statistical error.
Plots of the statistical convergence of the other moments can be found in Figures \ref{fig:RTI_convergence_D10} and \ref{fig:RTI_convergence_D20} in Appendix \ref{RT spatial eddy diffusivity moments}.

\begin{figure}
    \centering
    \begin{subfigure}[]{0.49\textwidth}
        \includegraphics[width=\textwidth]{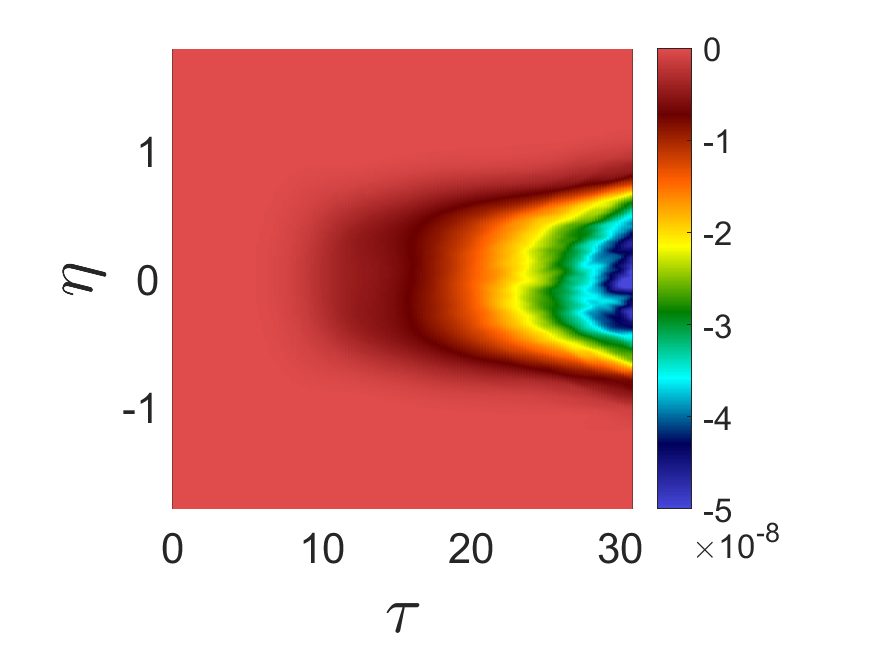}
        \subcaption[]{Standard method \edit{with separate donors}, $1,000$ realizations.}
    \end{subfigure}
    \begin{subfigure}[]{0.49\textwidth}
        \includegraphics[width=\textwidth]{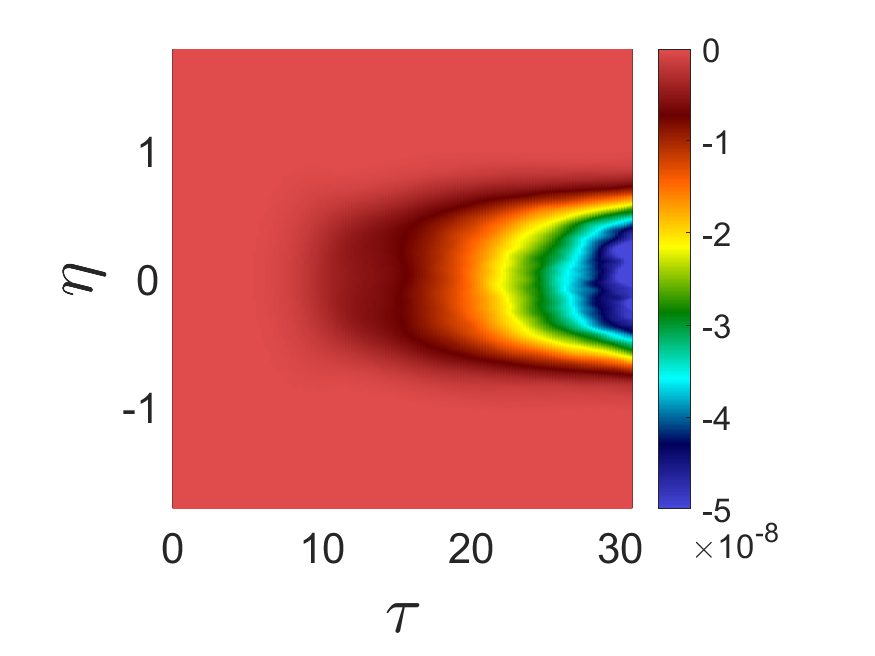}
        \subcaption[]{Decomposition method \edit{with a single donor}, $100$ realizations.}
    \end{subfigure}
    \caption{Qualitatively-similar states of statistically convergence of $D^{01}$ (normalized by $t$) for each method. }
    \label{fig:RTI_moments_converged}
\end{figure}

Figure \ref{fig:RTI_moments_converged} shows plots of $D^{10}$ measurements with qualitatively similar levels of statistical error.
Statistical convergence is achieved with a higher number of realizations for \edit{standard MFM with separate donors} ($1,000$) compared to \edit{decomposition MFM} ($100$).
This suggests that the decomposition MFM\edit{, because it uses a single donor simulation,} may offer a speedup factor of about ten for statistical convergence \edit{over MFM using separate donors}.

\subsection{Impact on constructing closure operators}

Statistical convergence is crucial for development of accurate closure models.
High statistical error can obstruct analysis by causing incorrect conclusions about the convergence of closure model predictions.
Here, we demonstrate the impact of statistical error on the matched moment inverse (MMI) procedure for constructing a closure operator for 2D RT instability.
MMI is a systematic method for modeling the nonlocal eddy diffusivity based on MFM-measured eddy diffusivity moments \citep{liu2023systematic}.
When applied to this spatiotemporal 2D RT problem, the result of the method is an inverse operator in the following form:
\begin{align}
    \left[1+a^{01}\frac{\partial}{\partial t}+a^{10}\frac{\partial}{\partial x_2}+a^{20}\frac{\partial^2}{\partial x_2^2}{+\dots}\right](-\langle u_2'c' \rangle )=a^{00}\frac{\partial \langle c \rangle }{\partial x_2},
\label{eq:MMI_RTI}
\end{align}
where $a^{mn}(x_2,t)$ are model coefficients determined using a process detailed in \citet{liu2023systematic} and \citet{lavacot2024}.
We use this form instead of truncating the Kramers--Moyal expansion in Equation \eqref{eq:RTI_moments} for several reasons. 
First, it is well known in the literature that the Kramers--Moyal expansion does not converge.
That is, finite truncation of the expansion leads to divergent results.
This property of the Kramers--Moyal expansion was proven by \citet{pawula1967approximation} and was shown to be true in modeling the eddy diffusivity by \citet{liu2023systematic} and specifically in 2D RT instability by \citet{lavacot2024}.
Secondly, a model involving truncation of the Kramers--Moyal expansion would be challenging to implement numerically in this spatiotemporal problem, since this would require time advancing spatial gradients of mixed derivatives.
The inverse operator in Equation \eqref{eq:MMI_RTI} addresses both of these issues.
Once the coefficient fields $a^{mn}$ are determined using the MMI procedure, Equation \eqref{eq:MMI_RTI} can be directly time-integrated using a partial differential equation solver.
Details on this and the MMI procedure can be found in \citet{liu2023systematic}.

In \citet{lavacot2024}, the importance of eddy diffusivity moments for modeling mean scalar transport was investigated using the standard MFM \edit{with separate donors}.
Different combinations of moments (i.e., different truncations of terms in the MMI operator in Equation \eqref{eq:MMI_RTI}) were tested to assess the importance of each moment.
Here, we examine the truncation of the MMI operator to the four terms shown in Equation \eqref{eq:MMI_RTI}.
Construction of this model form requires the eddy diffusivity moments $D^{00}$, $D^{10}$, $D^{01}$, and $D^{20}$.
We compare results of models constructed directly using measurements of $D^{mn}$ from the standard MFM \edit{with separate donors} and the decomposition MFM. 
For both methods, we use $D^{mn}$ measurements averaged over $200$ realizations.
At this number of realizations, the moments are visually converged for the decomposition MFM, but not the standard MFM \edit{using separate donors}.

\begin{figure}
    \centering
    \begin{subfigure}[]{0.49\textwidth}\includegraphics[width=0.9\textwidth]{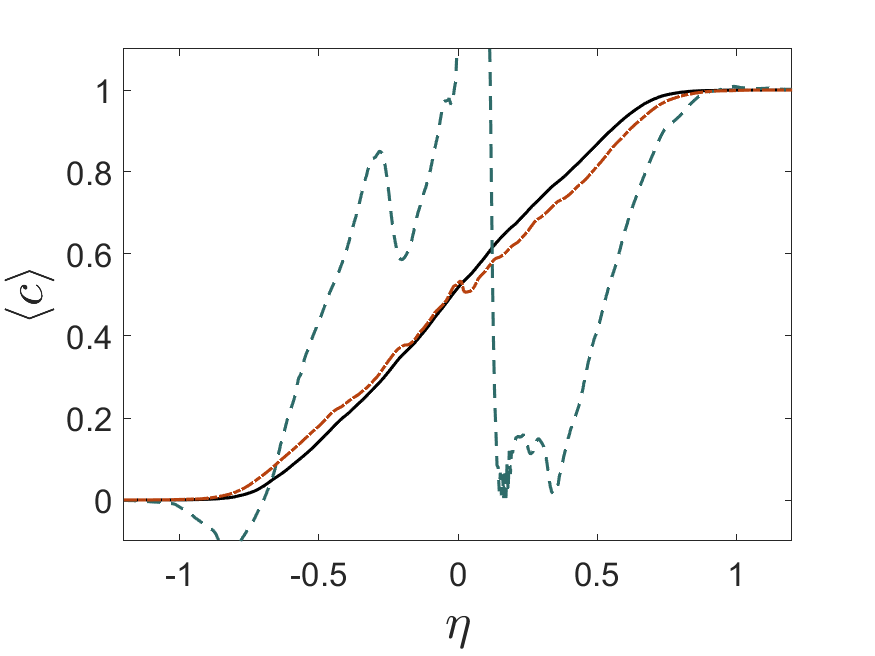}
        \subcaption[]{}
    \end{subfigure}
    \centering
    \begin{subfigure}[]{0.49\textwidth}\includegraphics[width=0.9\textwidth]{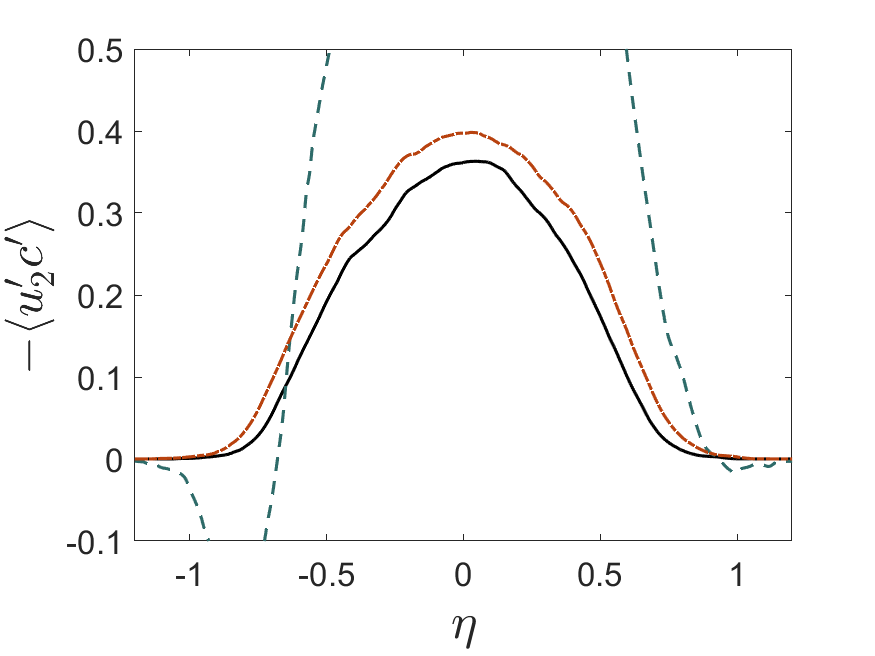}
        \subcaption[]{}
    \end{subfigure}
    \caption{(a) Mean concentration profiles and (b) turbulent scalar flux from DNS (solid black) and models using eddy diffusivity moments measured using the standard MFM \edit{with separate donors} (dashed green) and decomposition MFM (dash-dotted red). \edit{Eddy diffusivity moments} from both methods are averaged over $200$ realizations.}
    \label{fig:model_RTI}
\end{figure}

Figure \ref{fig:model_RTI} shows mean concentration and turbulent scalar flux profiles resulting from each of the models.
The results from the model using \edit{the separate-donor MFM} eddy diffusivity moments diverge significantly from the DNS results.
This may lead to the incorrect conclusion that addition of terms in the MMI operator does not lead to convergence in this case.
However, the source of this error is the large amount of statistical error in the higher-order moments.
On the other hand, the results from the model using the decomposition MFM moments agree much better with the DNS.
There appears to be some statistical error still at this number of realizations but \edit{significantly less than} with the MFM \edit{with separate donors}.
This highlights the importance of statistically-converged higher order moments in modeling.
Compared to the standard MFM \edit{using separate donors}, the decomposition MFM is a more efficient method for obtaining statistically-converged moments that are usable for constructing models.

\section{Conclusion}
In this work, modifications to the standard MFM are presented \edit{for enhanced statistical convergence}. 
We first demonstrate the utility of using a single donor simulation for the receiver equations.
Using a single donor prevents pileup of statistical error that may arise in separate donors, even when the same numerical initial conditions are used.
Due to the potentially slow statistical convergence of MFM using separate donors, we recommend MFM using only one donor, though that may not be the natural implementation in some codes.
\editt{
In cases where the repeated numerical solutions of the donor equations  produce identical instantaneous results (see Appendix \ref{channel flow}), using a single donor is not strictly necessary; however, it does have the added cost-saving benefit of not solving the donor equations multiple times. Moreover, for cases where the MFM forcing may be incompatible with the boundary conditions of the problem, decomposition MFM must be used which would also require using a single donor. 
}

\edit{
While the statistical error issue presented in this work is due to computational processes specific to the Ares code, sources of error between separate donors are not limited to computation. 
For example, when performing MFM for chaotic flows requiring multiple realizations for statistical convergence, one may consider computing new higher-order moments using a new donor but keeping previously-computed lower-order moments from a separate donor.
In this case, there are certainly differences between the donors of the lower-order and higher-order moments.
As we have shown in this work, small differences between separate donors significantly affect statistical error, and one should expect to incur larger statistical error in the higher-order moments.
Another interpretation of our results is that since differences in ensembles can result in amplified statistical error in the moments, simulations belonging to one ensemble should remain with that ensemble, i.e., they should not be used for statistics in a different ensemble.
Overall, this work highlights the nuances with the donor equations that one should consider to avoid additional statistical error when using MFM to compute the eddy diffusivity moments.
}

We also present decomposition MFM, \edit{which naturally uses a single donor}, for both scalar and momentum transport. For example, for scalar transport, instead of solving for the scalar $c$ as in standard MFM, decomposition MFM solves for a variable based on the scalar fluctuation $\cmom$, which is then expanded using the Kramers-Moyal expansion, allowing the mean forcing in the receiver equations to be handled semi-analytically. 
\editt{Decomposition MFM allows for consistent boundary condition treatment, such as in problems with periodic domains.}

Decomposition MFM differs from other methods for accelerating MFM in its purpose for measuring eddy diffusivity (or eddy viscosity) moments. Fast MFM~\cite{bryngelson2024fast} was developed for approximating the nonlocal and anisotropic eddy diffusivity for the entire domain by leveraging hidden sparsity in the discretized eddy diffusivity. Likewise, adjoint MFM~\cite{liu2024adjoint} was developed for economical targeted computation of the exact nonlocal and anisotropic eddy diffusivity at specific locations in the domain using an adjoint-based approach rather than eddy diffusivity moments. 

To demonstrate its utility, we apply decomposition MFM to 2D RT instability.
In this case study, we demonstrate computational savings of at least an order of magnitude in reaching statistical convergence of eddy diffusivity moments compared to MFM using separate donors.
\edit{This improved statistical convergence is since decomposition MFM inherently uses a single donor rather than separate donors.}
We show that this improved statistical convergence is substantial and significantly impacts analysis of closure operators, as poorly-converged eddy diffusivity moments can lead to incorrect conclusions.

\textbf{Acknowledgements.} 
This work was performed under the auspices of the US Department of Energy by Lawrence Livermore National Laboratory under Contract No. DE-AC52-07NA27344.
D.L. was additionally supported by the Charles H. Kruger Stanford Graduate Fellowship.
\edit{
We are also thankful to Dr. Sho Takatori for discussions on MFM and pointing us to connections to Generalized Taylor Dispersion Theory in the work of Dr. Howard Brenner.
}

\appendix
\section{Error analysis}
\label{sec:error_analysis}
We present the following error analysis to illustrate the propagation of error due to \edit{separate} donor simulations.
This analysis is done in one dimension ($x_1$) for simplicity.
First, we define $F^i$ to be the measurements of the turbulent scalar flux used to determine $D^i$:
\begin{align}
F^0 = \ens{-u_1'c'}|_{\frac{\partial\ens{c}}{\partial x_1}=1},\\
F^1 = \ens{-u_1'c'}|_{\frac{\partial\ens{c}}{\partial x_1}=x_1}. 
\end{align}
Since the numerical mean uses a finite number of ensembles, there exists statistical error when making the measurements $F^i$.
Examination of Equation \eqref{eq:f expansion} reveals that the statistical error arises due to $\cmom^{mn}$; $\frac{\partial\ens{c}}{\partial x_j}$ are deterministic as they are set by the macroscopic forcing.
In this manner, we can rewrite Equation \eqref{eq:f00 eqn} with statistical error:
\begin{align}
    c' + \varepsilon = \left[(\cmom^{0} + \varepsilon^0)  
    + (\cmom^{1} + \varepsilon^1)\frac{\partial}{\partial x_1} 
    + (\cmom^{2} + \varepsilon^2)\frac{\partial^2}{\partial x_1^2} 
    + 
    \dots \right] \frac{\partial \ens{c}}{\partial x_1}.
    \label{eq:f expansion w error}
\end{align}
Thus, statistical error can also be written as a Kramers--Moyal expansion:
\begin{align}
    \varepsilon = \left[ \varepsilon^0 
    + \varepsilon^1\frac{\partial}{\partial x_1} 
    + \varepsilon^2\frac{\partial^2}{\partial x_1^2} 
    + 
    \dots \right] \frac{\partial \ens{c}}{\partial x_1}.
    \label{eq:error_expansion}
\end{align}

In the following analysis, we use the notation $\varepsilon^{ij}$, where $i$ denotes the error associated with $\cmom^i$, as in Equation \eqref{eq:error_expansion}, and $j$ denotes the simulation used to determine $D^j$.
The addition of index $j$ is introduced, because when \edit{separate} donor simulations are used in MFM, each simulation has its own statistical error.
For example, $\varepsilon^{00}$ is not necessarily the same as $\varepsilon^{01}$, despite them both being errors associated with $\cmom^0$.
The measurement of $D^0$ can then be written as
\begin{align}
    D^0 = F^0 + \varepsilon^{00},
\end{align}
where $\varepsilon^{00}$ arises from substituting $\frac{\partial\ens{c}}{\partial x_1}=1$ into Equation \eqref{eq:error_expansion}.
Similarly, $D^1$ can be written as
\begin{align}
    D^1 = F^1 - x_1D^0 + \varepsilon^{01}x_1 + \varepsilon^{11}
    = F^1 - x_1F^0 + x_1(\varepsilon^{01}-\varepsilon^{00}) + \varepsilon^{11}
\end{align}
If one donor simulation is used for all receiver equations in MFM, the third term above disappears, since $\varepsilon^{00}=\varepsilon^{01}=\varepsilon^{0}$.
When \edit{separate} donors are used, $\varepsilon^{00}$ and $\varepsilon^{01}$ are not the same, so the overall statistical error scales with $x_1$, making statistical convergence for $D^1$ slower than for $D^0$.
While this analysis is presented in one dimension, the variable $x_1$, can be considered as either a spatial or temporal variable.
Since time can become large in numerical simulations, and measurements are often taken in late time, the statistical convergence of the first temporal moment is especially slow.

This analysis can be extended to higher-order moments.
For example, we analyze the error propagation in computing $D^2$:
\begin{align}
    D^2 &= F^2 - x_1D^1 - \frac{x^2}{2}D^0 + \varepsilon^{22}
    \\
    &= F^2 - x_1F^1 - \frac{x_1^2}{2}F^0 
    + x_1^2(\varepsilon^{10}-\varepsilon^{00})
    + \frac{x_1^2}{2}(\varepsilon^{00}-\varepsilon^{02}) 
    + x_1(\varepsilon^{12}-\varepsilon^{11})
    + \varepsilon^{22}.
\end{align}
If one donor is used, the fourth, fifth, and sixth terms vanish.
If \edit{separate} donors are used, those terms remain and the overall statistical error scales by $x_1^2$, resulting in even slower statistical convergence for $D^2$.

It must be noted that this analysis assumes $D^0$ is constant in time, which is not true for unsteady flows such as RT.
Additionally, we do not predict the scalings of the $\varepsilon^{ij}$ with space or time.
This is why the statistical error plots in Figure \ref{fig:stat_error} do not exhibit the exact scalings derived here.
The goal of this analysis is not to provide the scalings but to illustrate the error amplification in higher-order moments and the presence of this extra error in MFM simulations using separate donors.


\section{Turbulent channel flow case study: Effect of using separate donors vs.\ a single donor and compatibility with decomposition MFM}
\label{channel flow}

 \begin{figure}[h]
    \centering
    \begin{subfigure}[]{0.46\linewidth}
    \includegraphics[width=\linewidth]{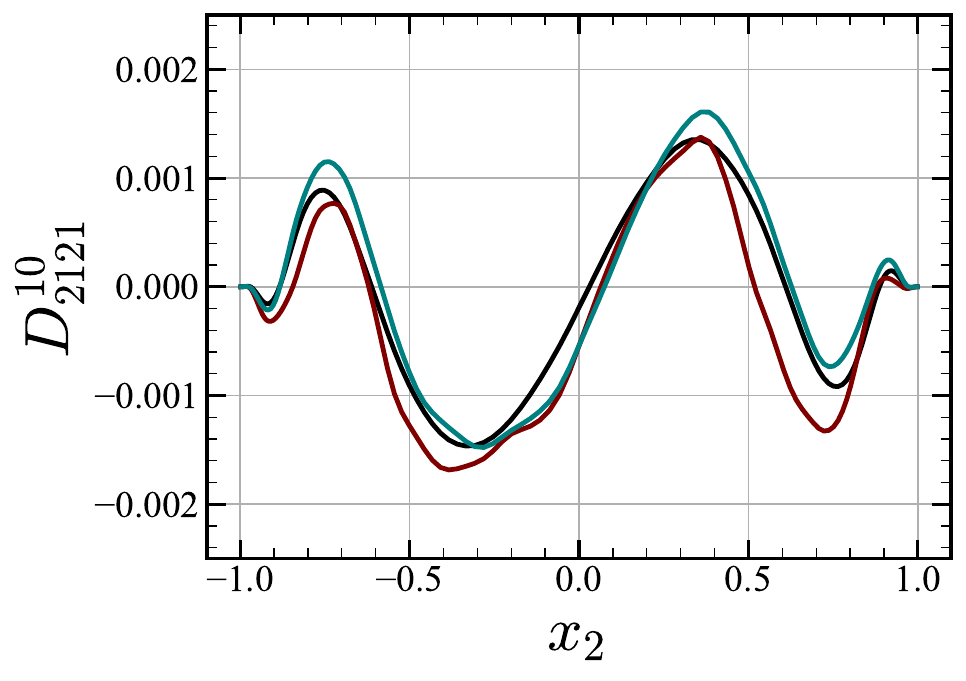}
    \subcaption[]{}
    \end{subfigure}
    \begin{subfigure}[]{0.461\linewidth}
    \includegraphics[width=\linewidth]{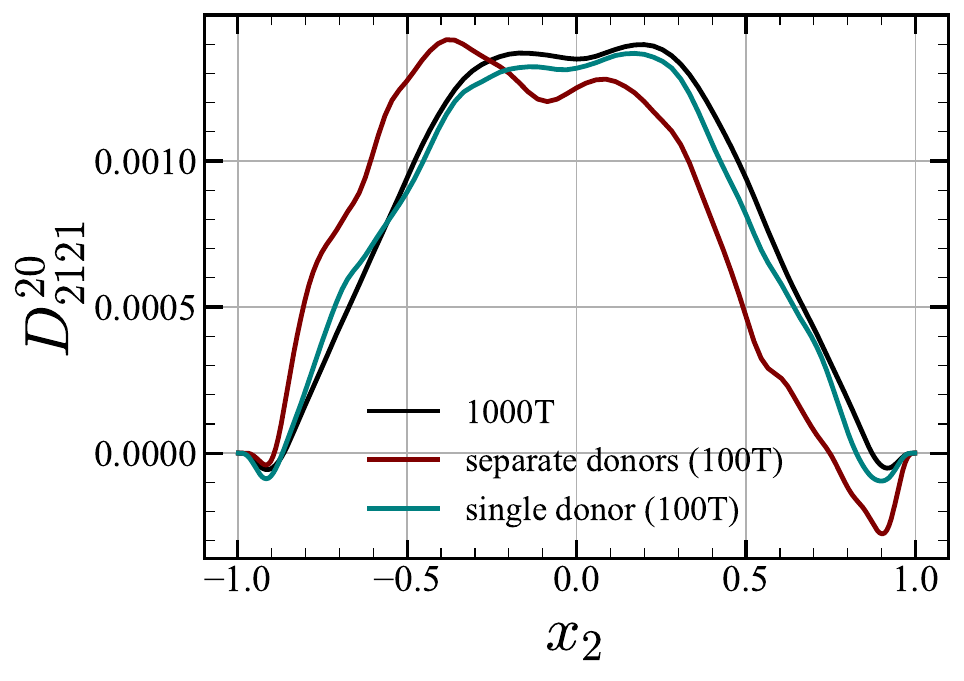}
    \subcaption[]{}
    \end{subfigure}
    \caption{\edit{Comparison of using standard MFM with separate donors vs.\ a single donor for quantifying the first and second spatial moments of the nonlocal eddy viscosity in turbulent channel flow at $Re_\tau=180$. The converged solution using standard MFM with a single donor at $1000$ eddy turnover time ($1000T$) is shown for comparison. Using separate donors clearly has more error at $100T$ than using a single donor.}}
    \label{fig:same_vs_diff}
\end{figure}

\begin{figure}[h]
    \centering
    \includegraphics[width=0.5\linewidth]{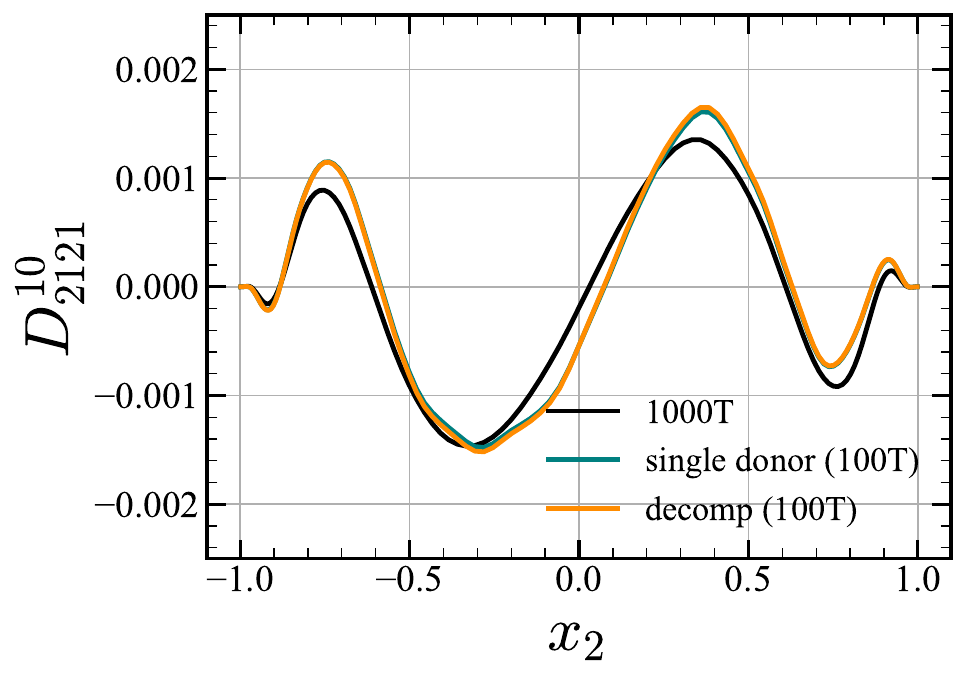}
    \caption{\edit{Comparison of using standard MFM with a single donor vs.\ decomposition MFM for quantifying the first spatial moment of the nonlocal eddy viscosity in turbulent channel flow at $Re_\tau=180$. Standard MFM with a single donor and decomposition MFM are almost identical at $100T$.}}
    \label{fig:standard_vs_decomp}
\end{figure}

In this appendix, we demonstrate the effect of using separate donors vs.\ a single donor in the context of turbulent channel flow. We then examine the effect of using decomposition MFM in conjuction with the single donor simulations.
This test case also serves as a demonstration of decomposition MFM for momentum as developed in Section \ref{sec:momentum}.

\editt{The turbulent channel flow does not suffer from the processor issues seen in the RT case, i.e., repeated numerical solutions of the turbulent channel flow donor equations with the same initial condition are identical and using separate donors would produce the same result as using a single donor. For the purposes of evaluating the equivalent effect of ``separate donors" vs.\ ``single donor," we introduce differences in the initial condition. ``Separate donors'' uses different initial conditions for the donor simulations whereas ``single donor'' uses the same initial condition for all donor simulations. While truly using a single donor, e.g., the implementation shown in Figure \ref{fig:MFM_diagram}b, is not strictly necessary for this case, it does have the cost-saving advantages of not repeatedly solving the donor equations. Moreover, using a single donor enables the use of decomposition MFM which is needed for quantifying eddy viscosity components related to the streamwise gradient of the mean velocity for which the periodic boundary conditions are incompatible with the MFM forcing  \cite{park2024direct, liu2024adjoint}.} 

\edit{Similar to the setup of \citet{park2024direct}, we use a friction Reynolds number of $Re_\tau=180$ where $\tau=u_\tau/h$ and where $u_\tau$ is the friction velocity and $h$ is the channel half-height. Averaging is taken over the homogeneous streamwise ($x_1$) and spanwise ($x_3$) directions, and the Reynolds stresses are only a function of the wall-normal ($x_2$) direction. Figure \ref{fig:same_vs_diff} shows a comparison of standard MFM using separate donors vs.\ a single donor for the first and second moments of the eddy viscosity in the wall-normal direction, $D^{10}_{2121}$ and  $D^{20}_{2121}$, respectively. The tensor subscripts correspond to the shear component of the Reynolds stress, $-\ens{u_2'v_1'}$, and the velocity gradient, $\partial \ens{v_1}/\partial x_2$, as shown in Equation \eqref{eq:kramers-moyal channel}. The solution at $1000T$ is taken to be the converged solution where $T$ is the eddy turnover time.} 
\edit{Figure \ref{fig:same_vs_diff} clearly shows the amplified errors associated with using separate donors.}

\edit{Figure \ref{fig:standard_vs_decomp} shows a comparison between standard MFM using a single donor and decomposition MFM. The momentum equations for decomposition MFM are detailed in Equations \eqref{eq:v00 momentum}-\eqref{eq:v10 mass}.  The two methods have almost identical solutions at $100T$, which is expected since standard MFM and decomposition MFM are mathematically equivalent. All improvement in statistical convergence is due to the usage of a single donor rather than separate donors.}

\section{Rayleigh--Taylor spatial eddy diffusivity moments}
\label{RT spatial eddy diffusivity moments}

\begin{figure}
    \centering
    \resizebox{.8\linewidth}{!}{\input{RTI_convergence_D10.tex}}
    \caption{Convergence of $D^{10}$ (normalized by $h$) measurement using \edit{standard MFM with separate donors} and decomposition \edit{MFM} for the RT instability case.}
    \label{fig:RTI_convergence_D10}
\end{figure}

\begin{figure}
    \centering
    \resizebox{.8\linewidth}{!}{\input{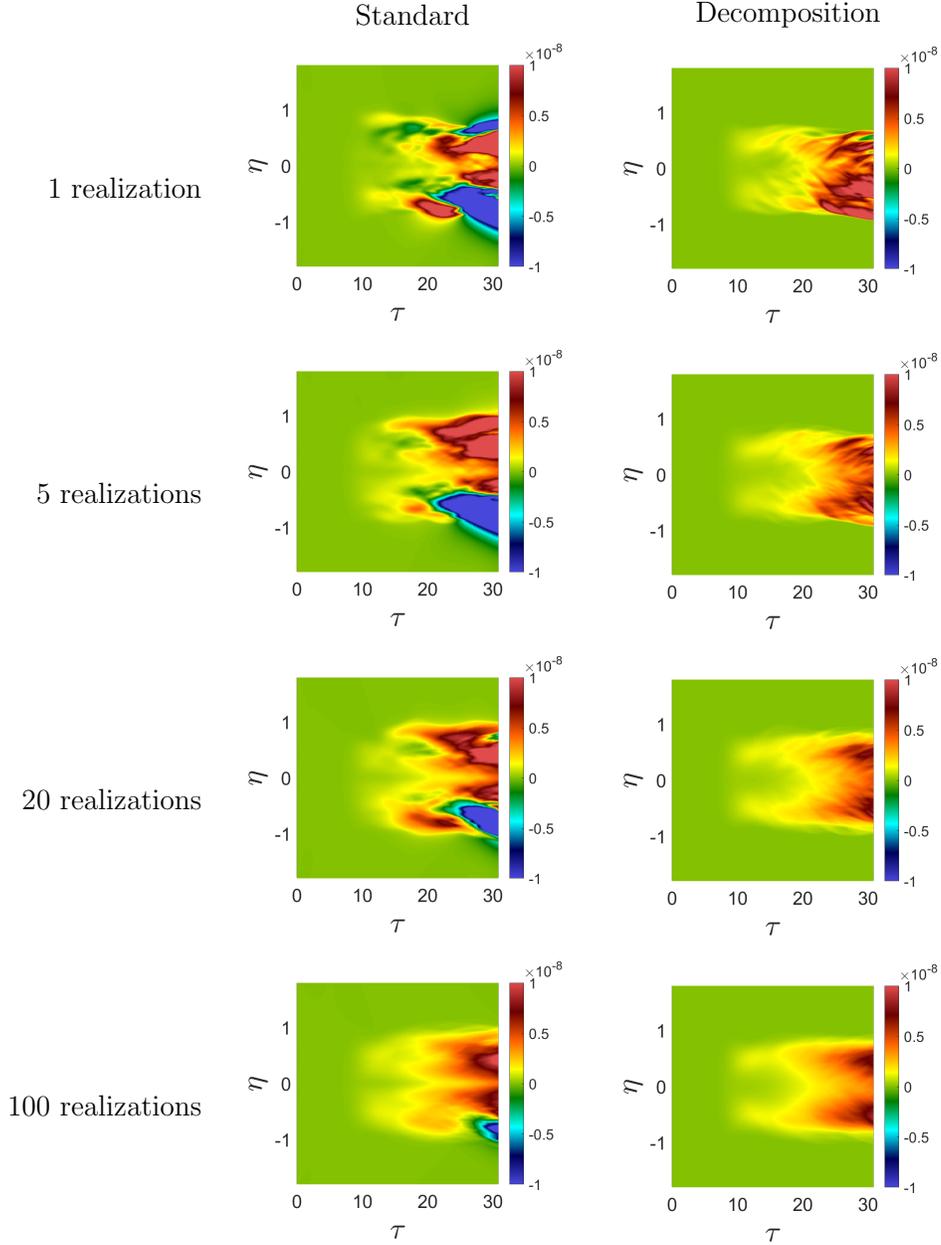}}
    \caption{Convergence of $D^{20}$ (normalized by $h^2$) measurement using \edit{standard MFM with separate donors} and decomposition \edit{MFM} for the RT instability case.}
    \label{fig:RTI_convergence_D20}
\end{figure}

\clearpage
\bibliographystyle{bibsty}
\bibliography{bibliography}

\end{document}